\documentclass[3p]{elsarticle}

\journal{Journal of Computational Physics}
\usepackage{lineno,hyperref}
\usepackage[title]{appendix}
\usepackage{amsmath}
\usepackage{array}
\usepackage{amssymb}
\usepackage{bm}
\usepackage{algorithm}
\usepackage{algpseudocode}
\usepackage{enumerate}
\usepackage{subcaption}
\usepackage{paralist}
\usepackage{multirow}

\begin{document}

\begin{frontmatter}

\title{Consistent, energy-conserving momentum transport for simulations of two-phase flows using the phase field equations}

\author[myaddress]{Shahab Mirjalili\corref{mycorrespondingauthor}}
\ead{ssmirjal@stanford.edu}
\cortext[mycorrespondingauthor]{Corresponding author}
\author[myaddress]{Ali Mani}
\ead{alimani@stanford.edu}
\address[myaddress]{Center for Turbulence Research, Stanford University, Stanford, CA 94305, USA}

\begin{abstract}
Realistic two-phase flow problems of interest often involve high $Re$ flows with high density ratios. Accurate and robust simulation of such problems requires special treatments. In this work, we present a consistent, energy-conserving momentum transport scheme in the context of a second order mass-conserving phase field method. This is achieved by (1) accounting for the mass flux associated with the right-hand-side of the phase field equation in the convective flux of the conservative form of the momentum transport equation---a correction absent in previous phase field simulations (2) utilization of non-dissipative spatial discretization. We demonstrate accuracy and robustness improvements from our proposed scheme via numerical tests, including a turbulent case of a water jet subject to air cross-flow. Our proposed modifications to the momentum transport equation can be quite readily extended to general conservative phase field methods. Additionally, in the framework of this phase field method, we present a free energy-based surface tension force calculation scheme. This scheme, which significantly reduces spurious currents, is based on a general paradigm that can also be extended to other two-phase flow methods.
\end{abstract}


\end{frontmatter}
\section{Introduction}
\label{sec:Intro}
In the past decades, various numerical methods have been proposed for simulating two-phase flows. Interface-capturing methods have been more popular due to the automatic treatment of topological changes. A review of the different classes of interface-capturing methods can be found in \cite{Mirjalili_ARB}. Phase field methods, which are also known as diffuse interface methods have emerged as promising approaches for simulating two-phase flows. In these methods, the transport equation governing the phase indicator is modified by incorporating physical effects that govern capillary interfaces. Although in realistic immiscible two-phase flows the physical thickness of the interface is virtually impossible to numerically resolve, these methods offer some desirable properties that have attracted the interest of two-phase flow modelers in recent years \citep{Anderson1998,Badalassi2003,Ding2007}. This can be primarily attributed to the following advantages:
\begin{itemize}
    \item Simplicity and ease of programming: Since phase advection is performed via discretization of partial differential equations (PDE's), such algorithms are much easier to implement compared to alternatives such as geometric VOF methods which require sophisticated computational geometry.
    \item Cost and scalability: The cost of time-integrating a PDE governing the evolution of the phase field is much lower than alternative options. Moreover, as opposed to levet-set and VOF schemes, phase field methods are blind to the interface location, rendering them automatically load-balanced.
    \item Smooth field: Similar to methods based on advection of a level-set function, phase field methods have the luxury of computing normal vectors and curvature values directly from a readily available smooth field. 
    \item Mass conservation: Phase field methods based on discretizing PDE's that are in conservative form conserve the mass of the system. Contrary to traditional level-set methods, since there is no reinitialization step in phase field methods, conservation of mass of the system is guaranteed.
\end{itemize}
  Traditionally, phase field methods have been based on the Cahn-Hilliard or Allen-Cahn equations, which are two important gradient flows of the Ginzburg-Landau-Wilson free energy functional. In a bounded physical domain given by $\Omega$, this energy functional is defined on the $H^1(\Omega)$ space in the form 
\begin{equation}
F:{ H }^{ 1 }(\Omega )\rightarrow [0,\infty ], F(\phi )=\frac { 1 }{ 2 } \int _{ \Omega  }{ { \epsilon  }^{ 2 }{ \left| \nabla \phi  \right|  }^{ 2 }dV }+ \int _{ \Omega  }{ W(\phi )dV }, 
\label{CH_energy_defn}
\end{equation}
where $\phi=-1$ and $\phi=+1$ represent the pure phases and $W(s)= (1-{ s }^{ 2 }) ^{ 2 }/4$ is the mathematically approximated double-well potential. In the absence of fluid flow, the steady state solution to the Cahn-Hilliard and Allen-Cahn equations minimizes the Ginzburg-Landau-Wilson free energy functional in two different norms. This enables these models to capture effects such as coarsening in binary-fluid mixtures and Ostwald ripening in systems in which phase change or phase separation can occur. However, in the two-phase flow community, phase field models are used in the same spirit as such sharp interface capturing schemes (VOF, level-set, etc.), with no reference to interface thermodynamics and continuum phase transitions, but instead solely employed for capturing the advection of phase interfaces due to fluid flow. 

Owing to its conservative form, the Cahn-Hilliard equation is a popular option within the two-phase flow community. This equation, given by
\begin{equation}
\frac { \partial \phi  }{ \partial t } +\nabla \cdot (\vec { u } \phi )=-{ \nabla  }^{ 2 }\left [{ \epsilon  }^{ 2 }{ \nabla  }^{ 2 }\phi -W'(\phi )\right ],
\label{Cahn_Hilliard}
\end{equation}
is the $H^{ -1 }(\Omega )$ gradient flow of the energy functional defined in Equation \ref{CH_energy_defn}. For this specific phase field method, \cite{Jacqmin1999} showed how surface tension force can be defined such that total energy (kinetic energy plus surface energy) is only dissipated, causing spurious currents to vanish. Most articles on Cahn-Hilliard, including \cite{Jacqmin1999}, have focused on equal or low density ratios. \cite{Ding2007} laid the foundation for applying these equations to flows with high density ratios. Later, \cite{Shen_Yang} extended the work by \cite{Jacqmin1999} to non-unity density ratios by elaborating on how the momentum equation for Allen-Cahn and Cahn-Hilliard systems should be modified such that these phase field systems admit discrete energy laws in which the total energy is non-increasing. \cite{Dong2012} later used this framework and presented a spectral element based solver which was suitable for handling the fourth order spatial derivatives in Equation \ref{Cahn_Hilliard}. 

Despite the Cahn-Hilliard model's advantage of having upper bounds on total energy which leads to robustness and stability, there are multiple issues with phase field methods based on the Cahn-Hilliard equation:
\begin{itemize}
    \item Artificial dissipation of total energy is undesirable, especially for realistic applications such as turbulent two-phase flows.
    \item Handling a fourth order spatial derivative is cumbersome.
    \item Equilibrium solutions can yield phase values outside of the $[1,-1]$ bounds away from the interfaces. For example, for a stationary spherical drop with radius $r$, the analytical equilibrium solution involves a far-field overshoot on the order of $\epsilon/r$\citep{Yue2007}. This shift in the equilibrium solution is problematic for high density ratios, as it can cause significant deviation in the density values, even resulting in regions of negative density. More importantly, since the Cahn-Hilliard equation is in conservative form, this results in artificial shrinkage (mass leakage) of drops and bubbles\citep{Yue2007}. 
    \item Coarsening effects are observed in two-phase flow simulations using Cahn-Hilliard equation. This is due to the fact that in order to minimize the energy functional in the domain, the right-hand side terms in the equation may cause spontaneous coalescence of drop/bubbles in a non-physical manner.
\end{itemize}
The third and fourth issues above are the most critical shortcomings of the original Cahn-Hilliard model. In order to keep the phase field function in the correct bounds, \cite{Dong2012} resorted to clipping out of bound values. This does not however counter the spontaneous shrinking (mass leakage) issue. Some methods such as those suggested by \cite{Wang2015,Li2016,Zhang2017} have attempted to reduce mass leakage by adding corrective terms to the right-hand side of Equation \ref{Cahn_Hilliard}. These "profile-corrected" and "flux-corrected" phase field methods result in reduction of spontaneous shrinkage (mass leakage) and coarsening effects\citep{Soligo2019}, but such fixes come with a penalty. Namely, the modified phase field equations in such methods are no longer gradient flows of an energy functional. As a result, converse to the original Cahn-Hilliard model, they do not admit discrete energy laws, undermining the rationale for solving a cumbersome fourth order PDE to capture two-phase interfaces. 

The Allen-Cahn equation is the ${ L }^{ 2 }(\Omega )$ gradient flow of the energy functional defined in Equation \ref{CH_energy_defn}, given by
\begin{equation}
\frac { \partial \phi  }{ \partial t } +\nabla \cdot (\vec { u } \phi )={ \epsilon  }^{ 2 }{ \nabla  }^{ 2 }\phi -W'(\phi ),
\label{Allen_Cahn}
\end{equation}
where $W'(\phi)=dW(\phi)/d\phi$. It is clear that the Allen-Cahn model is essentially a convection-diffusion equation with a source term. There is a wealth of numerical methods for solving such equations, and consequently, this easy to implement second order PDE is widely used in material science applications where phase change occurs. The downside to the Allen-Cahn model is that it is in nonconservative form. As such, it is not readily suitable for simulation of immiscible, incompressible two-phase flows with no phase change. Many authors have sought to rectify this issue by augmenting the original PDE with non-local corrections that allow for total mass conservation. Corrections using space and time dependent Lagrange multipliers have recently become popular for two-phase flow simulations using the Allen-Cahn equations \citep{Brassel2011,Kim2014,Zhai2015,Jeong2017,Joshi2018}. As an alternative approach, some researchers have modified the Allen-Cahn equation in a local sense. This has resulted in some new phase field models that do not suffer from the aforementioned intrinsic difficulties associated with Equations \ref{Cahn_Hilliard} and \ref{Allen_Cahn}, but instead combine their advantages.

\subsection{Conservative and bounded phase field method}
\label{sec:phase_field}
In \cite{Sun2007}, the curvature-driven flow in Allen-Cahn is subtracted out to obtain a second-order PDE suitable for two-phase simulations. Later on, \cite{Chiu_and_Lin} were inspired by the conservative-level-set literature to reformulate the phase field in \cite{Sun2007} in conservative form. We adopt the same PDE, given by
\begin{equation}
\frac { \partial \phi  }{ \partial t } +\nabla \cdot (\vec { u } \phi )=\gamma \nabla \cdot \left  [\epsilon \nabla \phi -\phi (1-\phi ) \left ( \frac { { \nabla  }\phi  }{ \left| { \nabla  }\phi  \right|  } \right ) \right ].
\label{phitrans}
\end{equation}
The right-hand side in Equation \ref{phitrans} is exactly the same as that used in the reinitialization step of the original conservative level-set method\citep{Olsson2005}. The authors in \cite{Chiu_and_Lin} time-integrated Equation \ref{phitrans} on a semi-staggered grid using a dispersion-relation-preserving upwind scheme specially developed for the convection-diffusion equation. Despite this, boundedness of $\phi$ was not guaranteed, and they had to resort to mass-redistribution and clipping to handle overshoots and undershoots. Such ad-hoc procedures can harm the accuracy and robustness of simulations, particularly at high $Re$ numbers and high density ratios.  

In \cite{Mirjalili_boundedness}, a simple non-dissipative space-time discretization for Equation \ref{phitrans} was introduced for staggered Cartesian grids. It was analytically and numerically shown that with appropriate selection of the free parameters ($\epsilon$ and $\gamma$), 
\begin{equation}
{ \epsilon  }/{ \Delta x }\ge\frac { { \gamma  }/{ { \left| \vec { u }  \right|  }_{ max } }+1 }{ 2{ \gamma  }/{ \left| \vec { u }  \right|  }_{ max } },
\label{general_crossover}
\end{equation}
explicit time-integration would always result in bounded values for $\phi$. We use this phase field approach in this work, which allows us to avoid diffusive upwinding, unphysical mass redistribution and interface reinitialization. In \cite{Mirjalili_comparison}, for an incompressible flow ($\nabla\cdot\vec{u}=0$), this interface capturing scheme was coupled to the non-conservative form of Navier-Stokes equation,
\begin{equation}
\frac{\partial \vec{u}}{\partial t}+\nabla\cdot(\vec{u}\otimes\vec{u})=\frac{1}{\rho}\left\{-\nabla P+\nabla\cdot[\mu(\nabla \vec{u}+\nabla^{T}\vec{u})]+\rho\vec{g}+\vec{F}_{ST}\right\},
\label{NS}
\end{equation}
where density and viscosity were linear with respect to $\phi$, and surface tension forces were computed via the the standard continuum surface force (CSF) method,
\begin{equation}
\vec{F}_{ST}=\sigma\kappa\nabla\phi.  
\end{equation}
In this formulation, $\kappa$, the curvature field, was computed using the normal vector, $\kappa=\nabla\cdot\vec{n}$ and the normal vector was given by $\vec{n}={\nabla\phi}/{\left|\nabla\phi\right|}$. All spatial derivatives in these equations were computed using central differences on a staggered grid. Namely, $P$,$\phi$,$\rho$ and $\mu$ were stored at cell centers while velocity values were stored at the center of their corresponding cell faces. This is a standard approach that results in desirable conservation properties for single phase flows when central difference schemes are used \citep{Morinishi1998}. For low $Re$ or low density ratio two-phase flows, \cite{Mirjalili_comparison} established the accuracy and efficiency of this fully coupled solver using several canonical two-phase flow tests in addition to a problem involving drop-pool impact.

\subsection{Consistent momentum transport}
\label{sec:consistent_conservative_momentum_transport}
Historically, in the framework of the one-fluid formulation, coupling Equation \ref{NS} to the interface-capturing step has been the most prevalent way of simulating incompressible two-phase flows \citep{bell1992,Tryggvason2011,Sussman2000,Desjardins2008,Kim2005,Ding2007,Dong2012,Chiu_and_Lin}. However, in various classes of two-phase flow methods, it has been observed that special numerical treatments must be performed to ensure numerical stability of this form of Navier-Stokes for high density ratios and/or high $Re$ numbers. In \cite{Sussman2007}, the authors were able to handle high density ratios while solving Equation \ref{NS} coupled to the CLSVOF method\citep{Sussman2000}. This was done by extrapolating the velocity of the denser phase into the lighter phase and using it to perform phase advection and calculation of the convective fluxes in the momentum equation. Other authors have used Equation \ref{NS} in conjunction with sharp interface methods by employing TVD time integration and upwinding schemes for spatial derivatives\citep{Fuster2009}. For diffuse interface methods, \cite{Ding2007} extended the work of \cite{Boyer2002} to solve Equation \ref{NS} coupled to the Cahn-Hilliard equation. The test cases in that work had limited $Re$, similar to the cases presented in \cite{Mirjalili_comparison}. In addition, this methodology did not satisfy any energy laws, unlike the seminal method of \cite{Jacqmin1999} which guarantees that the total energy of the system is non-increasing. This issue was resolved by \cite{Shen_Yang} by modifying the form of the momentum transport equation by introducing an auxiliary variable, $\sigma=\sqrt{\rho}$. This resulted in a method that is robust in simulating high density ratio two-phase flows with the Cahn-Hilliard phase field method. Similarly, \cite{Wang2019} proposed a stabilized phase field method using the entropy-viscosity method (EVM) that performs robustly at high $Re$ and high density ratios. Despite their stability, these proposed methodologies neither conserve momentum nor kinetic energy, even at the PDE level and in the absence of capillary and viscous forces. 

In \cite{Rudman1997}, the authors were able to successfully simulate flows with high density ratios by coupling their geometric VOF scheme to the conservative form of the Navier-Stokes equation. This form is given by:
\begin{equation}
\frac { \partial (\rho \vec{ u } ) }{ \partial t } +\nabla \cdot (\rho \vec{ u } \otimes \vec{ u } )=-\nabla P+\nabla \cdot \left[\mu (\nabla \vec { u } +\nabla ^{ T }\vec{ u } ) \right ]+{\vec{ F }}_{ ST }.
\label{momentum_generic_one_fluid_conservative}
\end{equation}
The main two features of their implementation are as follows:
\begin{itemize}
    \item Solving Equation \ref{momentum_generic_one_fluid_conservative}, where the convective term must be in conservative form.
    \item Using the mass flux that was computed in the phase advection step (geometrically in \cite{Rudman1997}) to convect velocity. 
\end{itemize}
Both of these features are critical to robustly simulating high density ratios. In the continuous limit of the equations, solving the conservative form of the momentum transport equation guarantees conservation of momentum and energy, albeit in the absence of capillary and viscous forces. Even in the presence of viscous and nonconservative capillary forces, this is expected to lead to improved robustness. Nevertheless, just solving the conservative form of the momentum equation does not yield a accurate/robust method for high density ratios. Rather, the second feature outlined above is a necessary condition for accurate and robust simulations of high density ratio two phase flows, especially at turbulent conditions. It is necessary to compute the convective flux of momentum in a manner consistent with the mass flux computed during phase advection. We will clarify this point further on for our diffuse interface method using theoretical analysis and numerical tests. While \cite{Rudman1997} employed a staggered grid configuration in addition to a finer mesh for mass advection, \cite{Bussmann2002} extended their methodology to unstructured grids. Later on, \cite{Raessi2012} applied their approach to level set methods. Using the same principals, it is commonly seen that state of the art two-phase flow methods, especially in the VOF class of schemes (where mass flux is readily available from the interface transport step) have adopted this consistent momentum transport approach\citep{Popinet2009,LeChenadec2013,Ivey_thesis,fuster2018}. Up until now, such mass-corrections in the momentum equation have been absent in phase field implementations.

As mentioned above, in diffuse interface methods for incompressible and immiscible two-phase flows, the nonconservative form of Navier-Stokes (Equation \ref{NS}) is typically solved. This has limited the application of these methods to either moderate density ratios or low $Re$ flows. Accordingly, for the diffuse interface model proposed in Section \ref{sec:consistent_conservative_momentum_transport}, through the numerical studies presented in \cite{Mirjalili_comparison} and \cite{Mirjalili_SNH}, we have observed robust and accurate simulations of low to moderate $Re$ two-phase flows with density ratios up to $1000$. However, in this work, through analysis and multiple numerical tests, including a simulation of a high density ratio and high $Re$ jet in cross-flow, we demonstrate that in order to solve high density ratio and/or high $Re$ two phase flows Equation \ref{NS} is not the appropriate equation to be coupled with the phase field evolution equation (Equation \ref{phitrans}). Instead, in the same spirit as \cite{Rudman1997}, it is necessary to solve a conservative form of the momentum equation while correcting the convective term to achieve consistent advection of mass and momentum. The main contribution of this work is the introduction of a consistent and kinetic-energy-conserving momentum transport methodology for conservative phase field equations. In the framework of the phase field equation given by Equation \ref{phitrans}, we show that our consistent and conservative momentum transport methodology can handle high density ratio turbulent two-phase flows, while showing the inaccuracies or lack of robustness if one uses \ref{NS} or \ref{momentum_generic_one_fluid_conservative} for momentum transport. It is important to point out that regardless of the density ratio, by using central difference operators on a staggered grid, our methodology discretely conserves mass, momentum and kinetic energy in the absence of capillary and viscous forces. To the best of our knowledge, among all the different methods belonging to the various classes of incompressible two-phase flow methods, our method is the first to accomplish this feat for non-unity density ratios.

In the following, in Section \ref{sec:ccmt_intro} we present our proposed form for the momentum transport PDE with appropriate physical justifications. Next, in Section \ref{sec:proof_KE_conservation} we provide analytical proof that in both continuous and discrete levels, the resulting methodology conserves kinetic energy in the absence of viscous and capillary forces. To show the improvement with the proposed modification, in addition to canonical tests, we present simulations of a practical jet in cross-flow involving turbulent conditions and high density ratios in Section \ref{sec:jet_crossflow}. These simulations are of note, because although many phase field methods have been proposed recently, realistic practical two-phase calculations, especially in turbulent conditions and realistic density ratios, are rarely carried out using phase field methods. Inspired by the work of \cite{Jacqmin1999}, we then develop a spurious current reducing, curvature-free surface tension force implementation for our novel phase field method in Section \ref{sec:EBCSF}. Finally, we conclude this article with a summary of our contributions in Section \ref{sec:Conclusions}. 

\section{Proposed consistent, kinetic-energy-conserving momentum transport equation}
\label{sec:ccmt_intro}
Let's consider a two-phase system, where the density of the two-phases are not equal. This is relevant to almost all practical two-phase flow systems. By considering the phase field model given by Equation \ref{phitrans} along with 
\begin{equation}
\rho=(\rho_{l}-\rho_{g})\phi+\rho_{g},
\label{rho}
\end{equation}
where $\rho_{l}$ and $\rho_{g}$ are the densities of the two fluids, the mass conservation equation is found to be
\begin{equation}
\frac { \partial \rho  }{ \partial t } +\nabla \cdot(\vec{u} \rho )=\nabla \cdot\vec{S}.
\label{mass_conservation_di}
\end{equation}
$\vec{S}$ is a conservative but artificial mass flux term given by
\begin{equation}
\vec{ S } =\gamma ({\rho}_{l}-{\rho}_{g}) \left[ \epsilon \nabla \phi -\phi(\phi - 1)\vec{n} \right ] =\gamma \left[ \epsilon \nabla \rho -\frac { ({ \rho  }_{ l }-\rho)(\rho -{ \rho  }_{ g }) }{ { \rho  }_{ l }-{ \rho  }_{ g } }\vec{n} \right ].
\label{S_defn}
\end{equation}
Here, $\vec{n}={\nabla\phi}/{|\nabla\phi|}$, as defined before. From Equation \ref{S_defn} it is clear that this mass flux is active only around the interface. This artificial flux, whose purpose is to maintain the hyperbolic tangent shape of the interface profile, displaces matter across the interface. The mere transport of matter via $\vec{S}$ from one side of the interface to another, results in a local increase/decrease in momentum and kinetic energy that is unaccounted for if one solves Equation \ref{NS} coupled to Equation \ref{phitrans}. This unaccounted momentum transfer can become catastrophic at high density ratios or even at moderate density ratios when the flow $Re$ is high. 

Our proposed solution is based on the notion that velocity field can be interpreted in two ways. First, velocity can be interpreted as the flux of volume (matter) per unit area per unit time. Secondly, velocity can also be considered as momentum per unit mass. For the transport of any quantity of interest, it is the first notion of velocity that must be used when computing the fluxes of that quantity. For instance, consider the problem of scalar field transport with a given velocity field using finite volume method. The flux of the scalar into a cell is computed via the velocity field which represents the volume flux into the cell. The concept of momentum is in fact irrelevant in such a problem. More precisely, it is the first notion of velocity, or more appropriately volumetrix flux, that transports any tensor of interest in Reynolds' transport theorem. 

Now, in the momentum equation for single phase constant density flows, the two notions of velocity are the same as $\rho$ is a constant. For two phase flows though, the transported quantity is $\vec{u}$, defined with the second notion as momentum per unit mass, while the mass flux that transports it must be computed in a manner which is discretely consistent with the discrete mass/phase advection equation. As mentioned above, this principle has been utilized in modern sharp-interface two-phase solvers\citep{Rudman1997,Bussmann2002,Raessi2012,LeChenadec2013,Ivey_thesis,fuster2018}. In our proposed diffuse interface framework, we are algebraically altering the mass fluxes via the right-hand side terms in Equation \ref{phitrans}. As such, the mass fluxes transporting any transportee, including momentum per unit mass, $\vec{u}$, are given by $\rho\vec{u}-\vec{S}$. With this, we propose the following modified consistent momentum equation, given by
\begin{equation}
\frac { \partial (\rho \vec{ u } ) }{ \partial t } +\nabla \cdot \left[(\rho \vec{ u }-\vec{S} )\otimes \vec{ u } \right ]=-\nabla P+\nabla \cdot \left[\mu (\nabla \vec{ u } +\nabla ^{ T }\vec{ u } ) \right ]+\vec{{ F }_{ ST }} .
\label{mom_con}
\end{equation}
This algebraic modification to the momentum transport equation is necessitated by the artificial terms present in the mass conservation equation. Discretely, the transporting mass flux in the convective term should be computed using exactly the same central difference and interpolation operators used in discretizing Equation \ref{phitrans} (see \cite{Mirjalili_boundedness}).

An important advantage of using Equation \ref{mom_con} for momentum transport is that in addition to momentum, in the absence of viscous and capillary effects, it conserves the kinetic energy of the system on the continuous and also discrete level by virtue of using central finite differences. Neither of the other forms of the momentum equation given by Equations \ref{NS} and \ref{momentum_generic_one_fluid_conservative} satisfy kinetic energy conservation even on the PDE level.

\subsection{Proof of kinetic energy conservation}
\label{sec:proof_KE_conservation}
Writing Equation \ref{mass_conservation_di} in index form, 
\begin{equation}
\frac { \partial \rho  }{ \partial t } +\frac { \partial { U }_{ i } }{ \partial { x }_{ i } } =0,
\label{mass_conservation_index_form}
\end{equation}
where ${U}_{i}=\rho{u}_{i}-{S}_{i}$ is the mass flux including the artificial contributions from the right hand side of Equation \ref{phitrans}. Now consider the index form of Equation \ref{mom_con} in the absence of viscous and capillary forces given by
\begin{equation}
    \frac { \partial \rho { u }_{ i } }{ \partial t } +\frac { \partial { U }_{ j }{ u }_{ i } }{ \partial { x }_{ j } } =-\frac { \partial P }{ \partial { x }_{ i } }.
    \label{mom_con_index_form}
\end{equation}
It is clear from the conservative form of the right hand side terms, that total momentum of the system is conserved. For a periodic domain $\Omega$ we have
\begin{equation}
    \frac { \partial  }{ \partial t } \int _{ \Omega  }{{ \rho  }{ u }_{ i }dV } =0.
    \label{total_mom_equation}
\end{equation}
For an incompressible flow, we have ${\partial {u}_{i}}/{\partial {x}_{i}}=0$. If we multiply Equation \ref{mass_conservation_index_form} by $-\frac{1}{2}{u}_{i}{u}_{i}$, and Equation \ref{mom_con_index_form} by ${u}_{i}$ (component by component), summation yields 
\begin{equation}
\frac { \partial  }{ \partial t } (\frac { 1 }{ 2 } { \rho  }{ u }_{ i }{ u }_{ i })=-\frac { 1 }{ 2 } { u }_{ i }{ u }_{ i }\frac { \partial { U }_{ j } }{ \partial { x }_{ j } } +{ u }_{ i }\frac { \partial { (U }_{ j }{ u }_{ i }) }{ \partial { x }_{ j } } -{ u }_{ i }\frac { \partial P }{ \partial { x }_{ i } } =-\frac { \partial { (U }_{ j }(\frac { 1 }{ 2 } { u }_{ i }{ u }_{ i })) }{ \partial { x }_{ j } } -\frac { \partial (P{ u }_{ j }) }{ \partial { x }_{ j } }. 
\label{KE_transport_continuous}
\end{equation}
In Equation \ref{KE_transport_continuous}, the two terms on the right hand side of the last equation are in conservative form and represent kinetic energy transport via the flux of mass and pressure work, respectively. These transport terms cannot contribute to the overall kinetic energy of the system. As such, for a periodic domain, $\Omega$, we have
\begin{equation}
    \frac { \partial  }{ \partial t } \int _{ \Omega  }{ (\frac { 1 }{ 2 } { \rho  }{ u }_{ i }{ u }_{ i })dV } =0.
    \label{total_KE_equation}
\end{equation}
Including the diffusive fluxes on the right hand side of Equation \ref{mom_con_index_form} would lead to a strictly negative term on the right hand side of Equation \ref{total_KE_equation}, representing dissipation of kinetic energy.

 We have so far shown that our proposed form for the momentum transport equation (Equation \ref{mom_con}) satisfies conservation of momentum and kinetic energy (Equations \ref{total_mom_equation} and \ref{total_KE_equation}). In comparison, Equation \ref{NS} neither conserves momentum nor kinetic energy, while Equation \ref{momentum_generic_one_fluid_conservative} is only momentum conserving. Now we focus on proving momentum and kinetic energy conservation in the discrete sense.

As mentioned before, we use staggered structured grids throughout this work. Velocity vectors and all fluxes, including the mass fluxes, are stored on their respective faces, while pressure, density, viscosity and the phase field variable are on cell centers. We carry out the proof in 2D, but extension to 3D is straight-forward. For this section, we concern ourselves with uniform Cartesian meshes. It is important to note that discrete conservation of kinetic energy is satisfied in the semi-discrete sense, where the temporal derivatives are not discretized. 

To simplify notation we denote the velocity in the $x$ and $y$ direction with $u$ and $v$. The x component of velocity on the left face of the $(i,j)-th$ cell is thus ${u}_{i-1/2,j}$. A similar format is used for the mass fluxes, by which $U$ and $V$ represent the mass flux in the $x$ and $y$ direction, respectively. Using central differences, the semi-discrete version of Equation \ref{mass_conservation_index_form} in 2D reads
\begin{equation}
    \frac { \partial { \rho  }_{ i,j } }{ \partial t } +\frac { { U }_{ i+1/2,j }-{ U }_{ i-1/2,j } }{ \Delta x } +\frac { { V }_{ i,j+1/2 }-{ V }_{ i,j-1/2 } }{ \Delta y } =0,
    \label{mass_conservation_semidiscrete}
\end{equation}
where $\Delta x$ and $\Delta y$ are the mesh size in the $x$ and $y$ directions. It is important to note that the mass fluxes must be discretely obtained from the corresponding fluxes in the phase advection routine, defined necessarily at the same location. In an incompressible flow, by solving a Poisson system we enforce that 
\begin{equation}
    \frac { { u }_{ i+1/2,j }-{ u }_{ i-1/2,j } }{ \Delta x } +\frac { { v }_{ i,j+1/2 }-{ v }_{ i,j-1/2 } }{ \Delta y } =0.
    \label{incompressible_semidiscrete}
\end{equation}
 The semi-discretized version of Equation \ref{mom_con_index_form} in the x direction reads
 \begin{multline}
     \frac { \partial (\frac { { \rho  }_{ i,j }+{ \rho  }_{ i-1,j } }{ 2 } { u }_{ i-1/2,j } )}{ \partial t } +\frac { \frac { { U }_{ i+1/2,j }+{ U }_{ i-1/2,j } }{ 2 } \frac { { u }_{ i+1/2,j }+{ u }_{ i-1/2,j } }{ 2 } -\frac { { U }_{ i-1/2,j }+{ U }_{ i-3/2,j } }{ 2 } \frac { { u }_{ i-1/2,j }+{ u }_{ i-3/2,j } }{ 2 }  }{ \Delta x } +\\ \frac { \frac { { V }_{ i,j+1/2 }+{ V }_{ i-1,j+1/2 } }{ 2 } \frac { { u }_{ i-1/2,j+1 }+{ u }_{ i-1/2,j } }{ 2 } -\frac { { V }_{ i,j-1/2 }+{ V }_{ i-1,j-1/2 } }{ 2 } \frac { { u }_{ i-1/2,j }+{ u }_{ i-1/2,j-1 } }{ 2 }  }{ \Delta y } +\frac { { P }_{ i,j }-{ P }_{ i-1,j } }{ \Delta x } =0.
     \label{mom_con_semidiscrete_x}
 \end{multline}
 A similar equation can be written for the $y$ component,
 \begin{multline}
     \frac { \partial (\frac { { \rho  }_{ i,j }+{ \rho  }_{ i,j-1 } }{ 2 } { v }_{ i,j-1/2 }) }{ \partial t } +\frac { \frac { { U }_{ i+1/2,j }+{ U }_{ i+1/2,j-1 } }{ 2 } \frac { { v }_{ i+1,j-1/2 }+{ v }_{ i,j-1/2 } }{ 2 } -\frac { { U }_{ i-1/2,j }+{ U }_{ i-1/2,j-1 } }{ 2 } \frac { { v }_{ i,j-1/2 }+{ v }_{ i-1,j-1/2 } }{ 2 }  }{ \Delta x } \\ \frac { \frac { { V }_{ i,j+1/2 }+{ V }_{ i,j-1/2 } }{ 2 } \frac { { v }_{ i,j+1/2 }+{ v }_{ i,j-1/2 } }{ 2 } -\frac { { V }_{ i,j-1/2 }+{ V }_{ i,j-3/2 } }{ 2 } \frac { { v }_{ i,j-1/2 }+{ v }_{ i,j-3/2 } }{ 2 }  }{ \Delta y } +\frac { { P }_{ i,j }-{ P }_{ i,j-1 } }{ \Delta y } =0.
     \label{mom_con_semidiscrete_y}
 \end{multline}
 It is once again clear from the conservative form of the fluxes, that momentum is discretely conserved for both $x$ and $y$ components. In other words, in a periodic domain
 \begin{equation}
     \frac { \partial  }{ \partial t } \sum _{ i,j }{ (\frac { { \rho  }_{ i,j }+{ \rho  }_{ i-1,j } }{ 2 } { u }_{ i-1/2,j }) } =\frac { \partial  }{ \partial t } \sum _{ i,j }{ (\frac { { \rho  }_{ i,j }+{ \rho  }_{ i,j-1 } }{ 2 } { v }_{ i,j-1/2 }) } =0,
     \label{total_mom_conservation_semidiscrete}
 \end{equation}
 where the location at which the momentum components are discretely defined is on the faces. In a similar manner we define the discrete kinetic energy of the system as 
 \begin{equation}
 KE=\sum _{ i,j }{\frac { 1 }{ 2 }  (\frac { { \rho  }_{ i,j }+{ \rho  }_{ i-1,j } }{ 2 } { u }_{ i-1/2,j }{ u }_{ i-1/2,j }+\frac { { \rho  }_{ i,j }+{ \rho  }_{ i,j-1 } }{ 2 } { v }_{ i,j-1/2 }{ v }_{ i,j-1/2 }) }.  
 \label{KE_discrete_defn}
 \end{equation}
 To find the evolution equation for ${\partial KE}/{\partial t}$, we interpolate Equation \ref{mass_conservation_semidiscrete} onto the $x$ and $y$ faces and multiply by $({-u}_{i-1/2,j}{u}_{i-1/2,j})/2$ and $({-v}_{i,j-1/2}{v}_{i,j-1/2})/2$, respectively. After summation, the result is added to the product of Equation \ref{mom_con_semidiscrete_x} with ${u}_{i-1/2,j}$ and the product of Equation \ref{mom_con_semidiscrete_y} with ${v}_{i,j-1/2}$. Finally, the result for all cells is summed to find the evolution of total kinetic energy. We assume that the domain has periodic boundary conditions on all boundaries. The temporal term is ${\partial KE}/{\partial t}$. The pressure terms become
 \begin{multline}
     \sum _{ i,j }{ \frac { { P }_{ i,j }{ u }_{ i-1/2,j }-{ P }_{ i-1,j }{ u }_{ i-1/2,j } }{ \Delta x } +\frac { { P }_{ i,j }{ v }_{ i,j-1/2 }-{ P }_{ i,j-1 }{ v }_{ i,j-1/2 } }{ \Delta y }  } =\\\sum _{ i,j }{ { P }_{ i,j }(\frac { { u }_{ i-1/2,j }-{ u }_{ i+1/2,j } }{ \Delta x } +\frac { { v }_{ i,j-1/2 }-{ v }_{ i,j+1/2 } }{ \Delta y } ) } =0,
     \label{pressure_terms_semidiscrete}
 \end{multline}
 where Equation \ref{incompressible_semidiscrete} is used to find the last equation. If we consider products in the form of ${U}_{a,b}{u}_{c,d}{u}_{e,f}$,
 \begin{multline}
   \sum _{ i,j }{ (\frac { -1 }{ 2 }  }{ u }_{ i-1/2,j }{ u }_{ i-1/2,j }\frac { { U }_{ i+1/2,j }-{ U }_{ i-3/2,j } }{ 2\Delta x } +\\{ u }_{ i-1/2,j }\frac { \frac { { U }_{ i+1/2,j }+{ U }_{ i-1/2,j } }{ 2 } \frac { { u }_{ i+1/2,j }+{ u }_{ i-1/2,j } }{ 2 } -\frac { { U }_{ i-1/2,j }+{ U }_{ i-3/2,j } }{ 2 } \frac { { u }_{ i-1/2,j }+{ u }_{ i-3/2,j } }{ 2 }  }{ \Delta x } )=\\ \sum _{ i,j }\frac { 1 }{ 4\Delta x } ({ U }_{ i+1/2,j }{ u }_{ i+1/2,j }{ u }_{ i-1/2,j }+{ U }_{ i-1/2,j }{ u }_{ i-1/2,j }{ u }_{ i+1/2,j }-\\{ U }_{ i-1/2,j }{ u }_{ i-1/2,j }{ u }_{ i-3/2,j }-{ U }_{ i-3/2,j }{ u }_{ i-3/2,j }{ u }_{ i-1/2,j })=0.  
     \label{Uu}
 \end{multline}
 And terms in the form of ${V}_{a,b}{u}_{c,d}{u}_{e,f}$ are
 \begin{multline}
     \sum _{ i,j }{ (\frac { -1 }{ 2 }  } { u }_{ i-1/2,j }{ u }_{ i-1/2,j }\frac { { U }_{ i+1/2,j }-{ U }_{ i-3/2,j } }{ 2\Delta x }\\ +{ u }_{ i-1/2,j }\frac { \frac { { V }_{ i,j+1/2 }+{ V }_{ i-1,j+1/2 } }{ 2 } \frac { { u }_{ i-1/2,j+1 }+{ u }_{ i-1/2,j } }{ 2 } -\frac { { V }_{ i,j-1/2 }+{ V }_{ i-1,j-1/2 } }{ 2 } \frac { { u }_{ i-1/2,j }+{ u }_{ i-1/2,j-1 } }{ 2 }  }{ \Delta y } )=\\ \sum _{ i,j } \frac { 1 }{ 4\Delta x } ({ V }_{ i,j+1/2 }{ u }_{ i-1/2,j+1 }{ u }_{ i-1/2,j }+V_{ i-1,j+1/2 }{ u }_{ i-1/2,j+1 }{ u }_{ i-1/2,j }-\\V_{ i,j-1/2 }{ u }_{ i-1/2,j }{ u }_{ i-1/2,j-1 }-{ V }_{ i-1,j-1/2 }{ u }_{ i-1/2,j }{ u }_{ i-1/2,j-1 })=0.
 \end{multline}
 Similarly, if we consider the summation of the terms in the form of ${U}_{a,b}{v}_{c,d}{v}_{e,f}$ and ${V}_{a,b}{v}_{c,d}{v}_{e,f}$ across all cells in the domain, they would both be zero. With that, we have shown that 
 \begin{equation}
     \frac{\partial KE}{\partial t}=\frac{\partial }{\partial t}\sum _{ i,j }{\frac { 1 }{ 2 } (\frac { { \rho  }_{ i,j }+{ \rho  }_{ i-1,j } }{ 2 } { u }_{ i-1/2,j }{ u }_{ i-1/2,j }+\frac { { \rho  }_{ i,j }+{ \rho  }_{ i,j-1 } }{ 2 } { v }_{ i,j-1/2 }{ v }_{ i,j-1/2 }) }=0,
     \label{KE_discrete_conservation}
 \end{equation}
 or in other words, Equation \ref{mom_con} coupled to Equation \ref{phitrans} (which is equivalent to Equation \ref{mass_conservation_di}) conserves the discrete kinetic energy of the system. Once again, including the discretized diffusive terms in the right hand side of Equations \ref{mom_con_semidiscrete_x} and \ref{mom_con_semidiscrete_y} results in ${\partial KE}/{\partial t}\le 0$.
 
 \section{Numerical tests}
 \label{sec:mom_con_numerical_tests}
In this section, we use two simple test cases in addition to simulations of a realistic jet in crossflow to demonstrate the necessity of using our proposed consistent and conservative version of the momentum transport equation (Equation \ref{mom_con}). 

\subsection{Case I}
\begin{figure} 
\centering
\begin{subfigure}{0.32 \linewidth}
\includegraphics[width=0.95\textwidth]{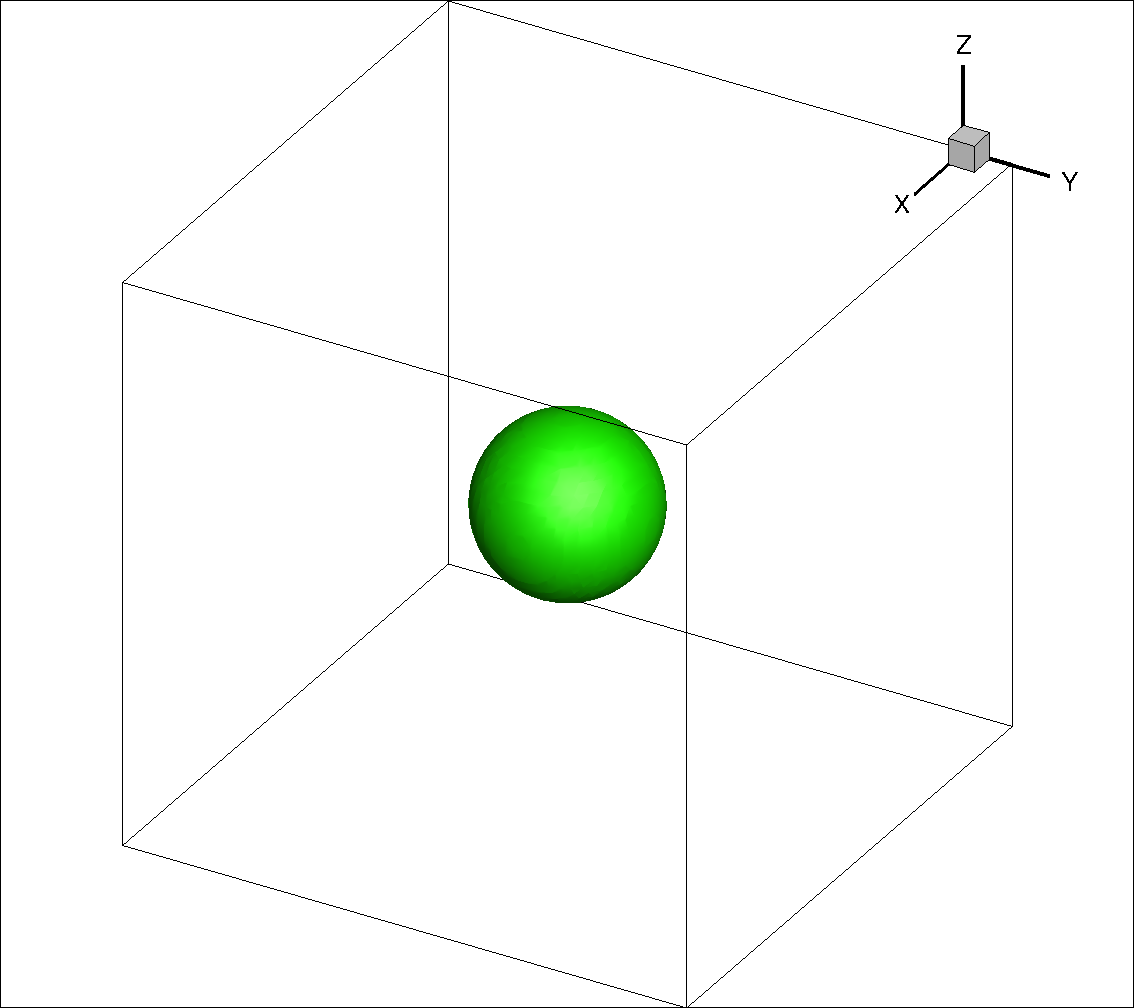} 
   \caption{}
   \label{fig:init_drop_advection_mom_cons} 
\end{subfigure}
\begin{subfigure}{0.32 \linewidth}
\includegraphics[width=0.95\textwidth]
{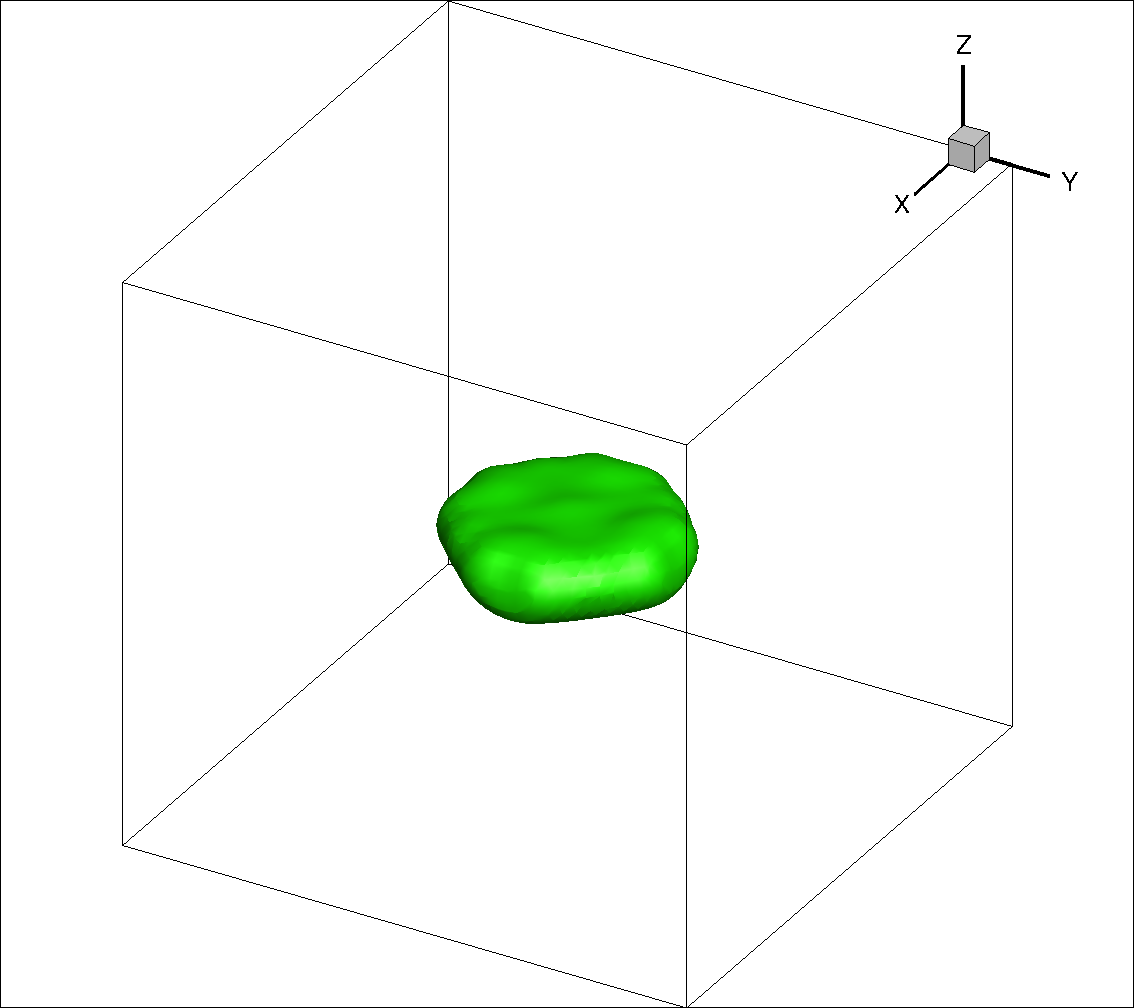} 
   \caption{}
  \label{fig:final_inconsist_drop_advection_mom_cons}  
   \end{subfigure}
   \begin{subfigure}{0.32 \linewidth}
\includegraphics[width=0.95\textwidth]
{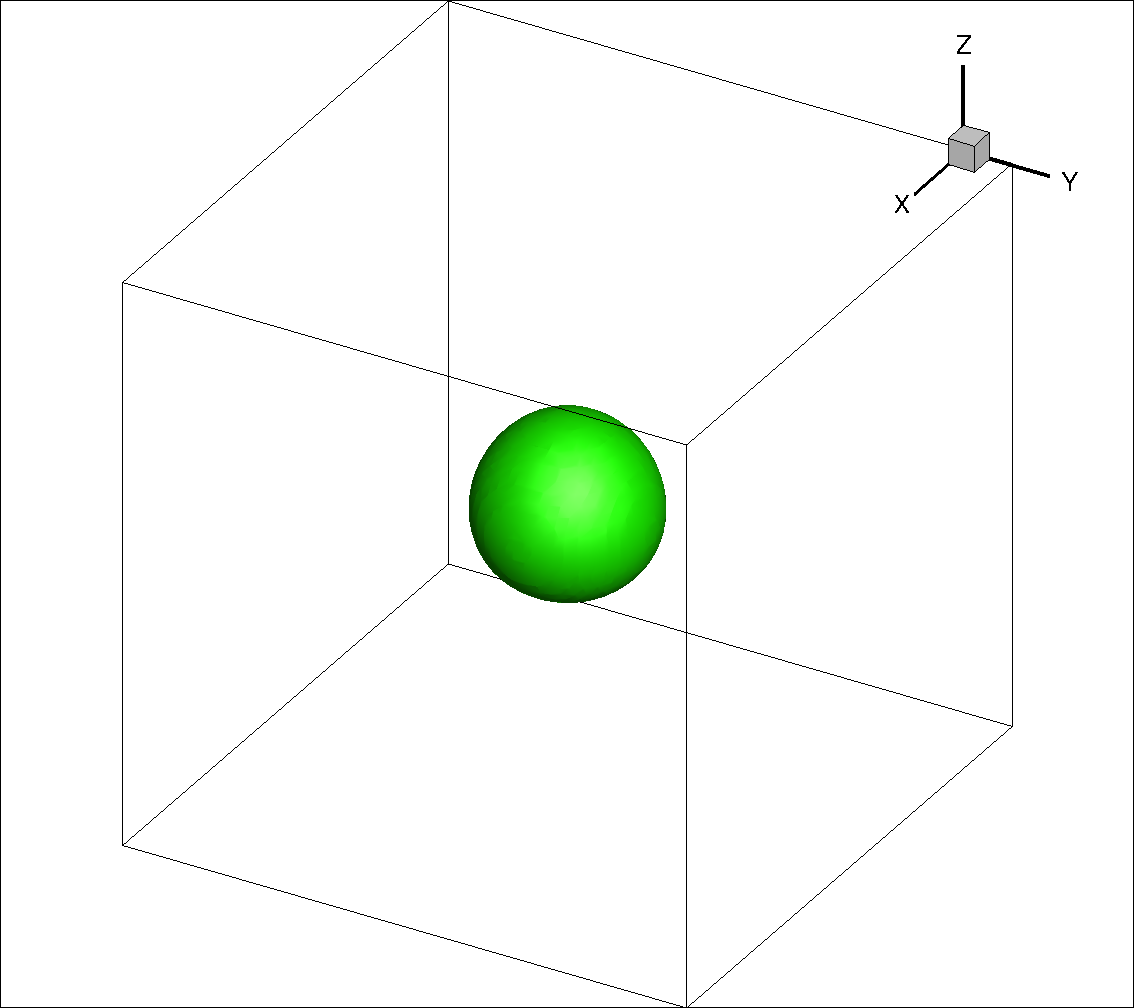} 
   \caption{}
  \label{fig:final_inconsist_drop_advection_mom_cons}  
   \end{subfigure}
    \caption{Drop interface profiles for drop advection with uniform velocity of $(0,0,1)$ in the domain (Case I) at (a) $t=0$, (b) $t=1$ using Equation \ref{mom_conservative_inconsistent}, (c) $t=1$ using Equation \ref{mom_con}}
       \label{fig:drop_advection_mom_cons_vs_incons}
\end{figure}

\begin{figure} 
\centering
\begin{subfigure}{0.48 \linewidth}
\includegraphics[width=0.95\textwidth]{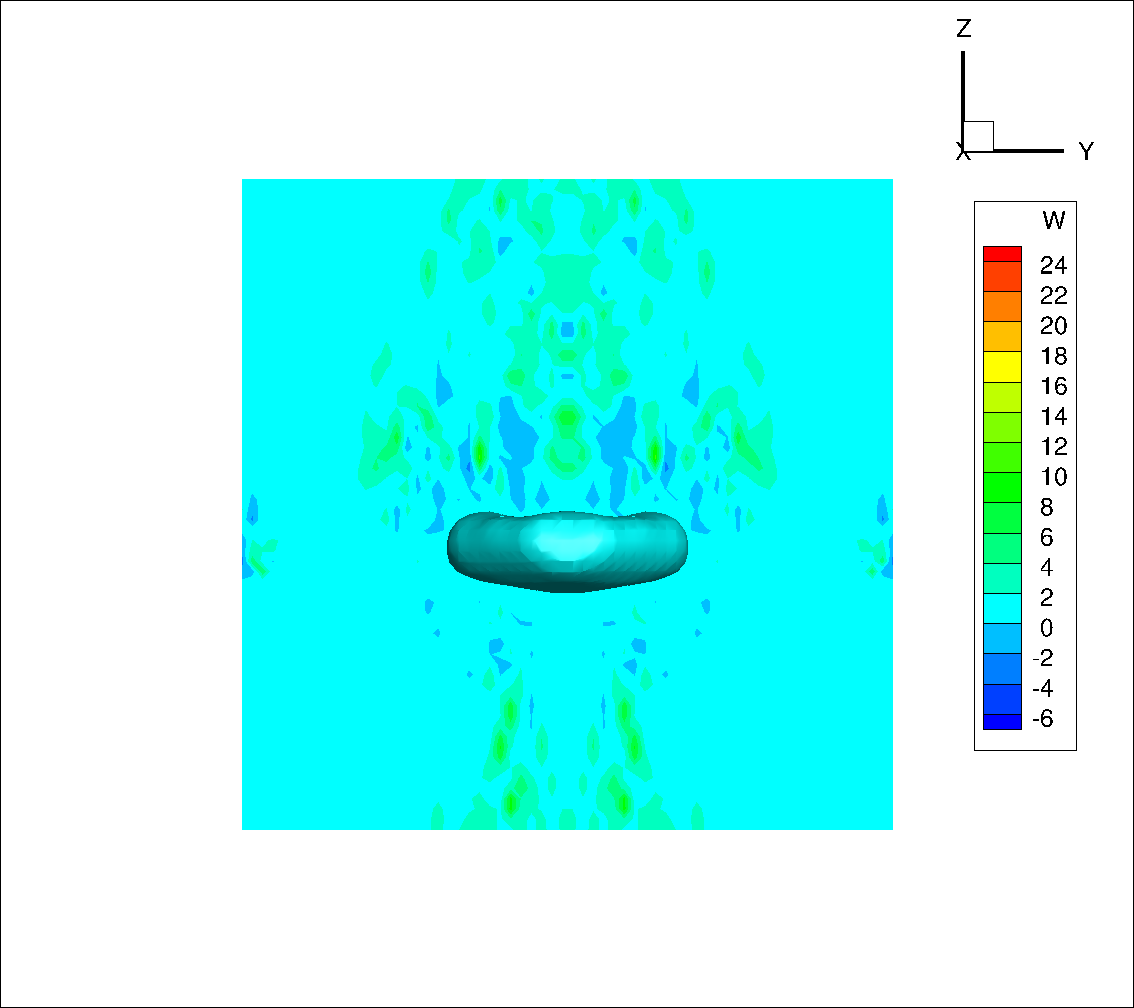} 
   \caption{}
   \label{fig:init_drop_advection_mom_cons} 
\end{subfigure}
\begin{subfigure}{0.48 \linewidth}
\includegraphics[width=0.95\textwidth]
{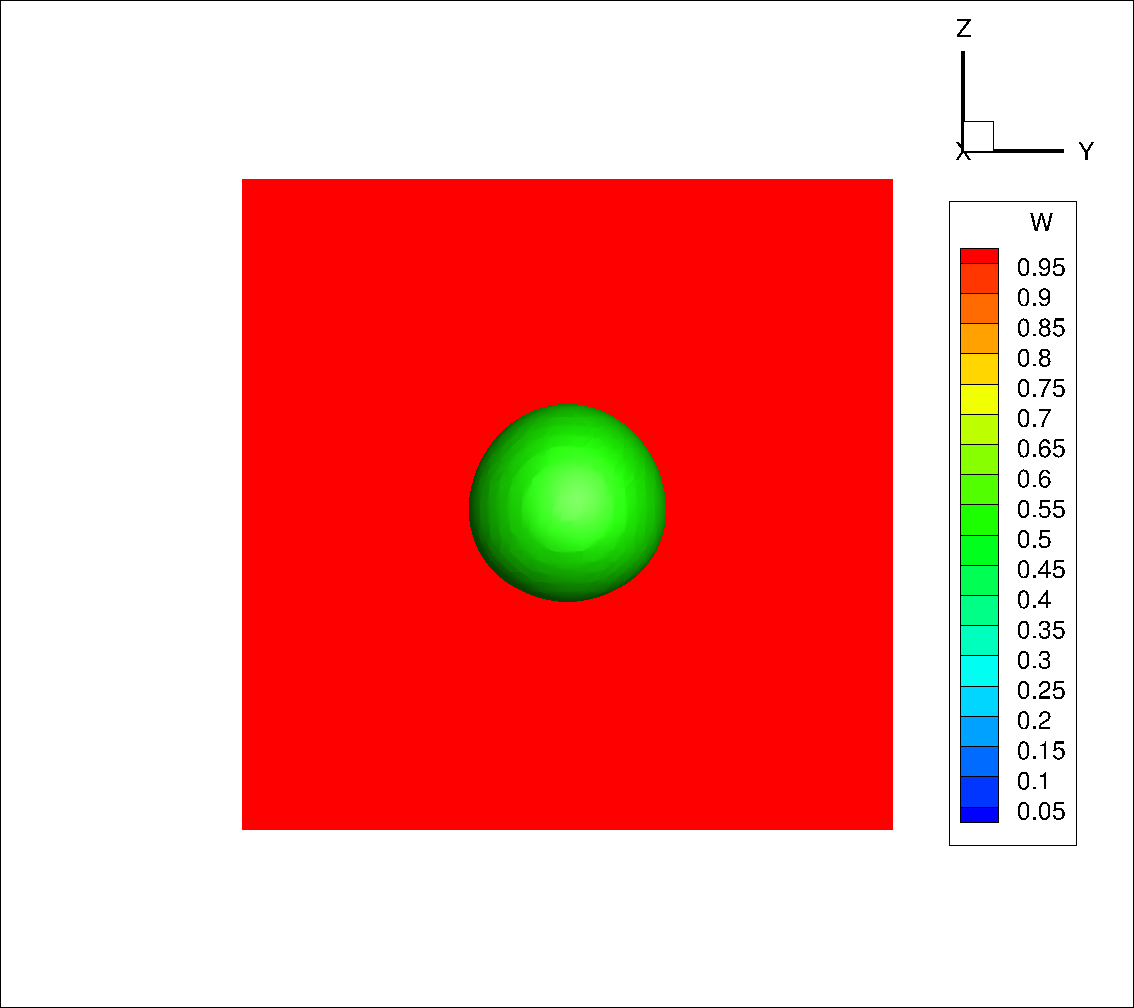} 
   \caption{}
  \label{fig:final_inconsist_drop_advection_mom_cons}  
   \end{subfigure}
   \begin{subfigure}{0.48 \linewidth}
\includegraphics[width=0.95\textwidth]
{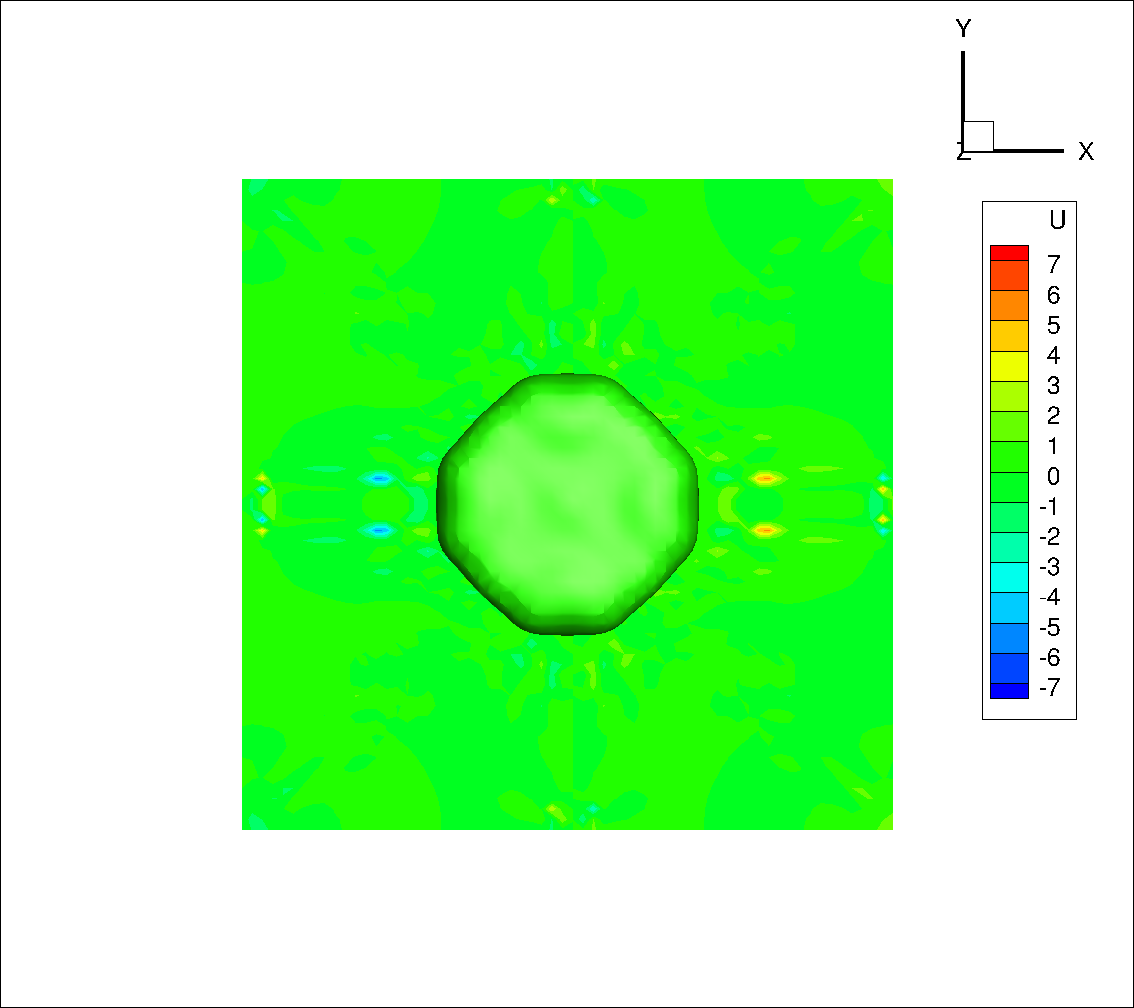} 
   \caption{}
  \label{fig:final_inconsist_drop_advection_mom_cons}  
   \end{subfigure}
      \begin{subfigure}{0.48 \linewidth}
\includegraphics[width=0.95\textwidth]
{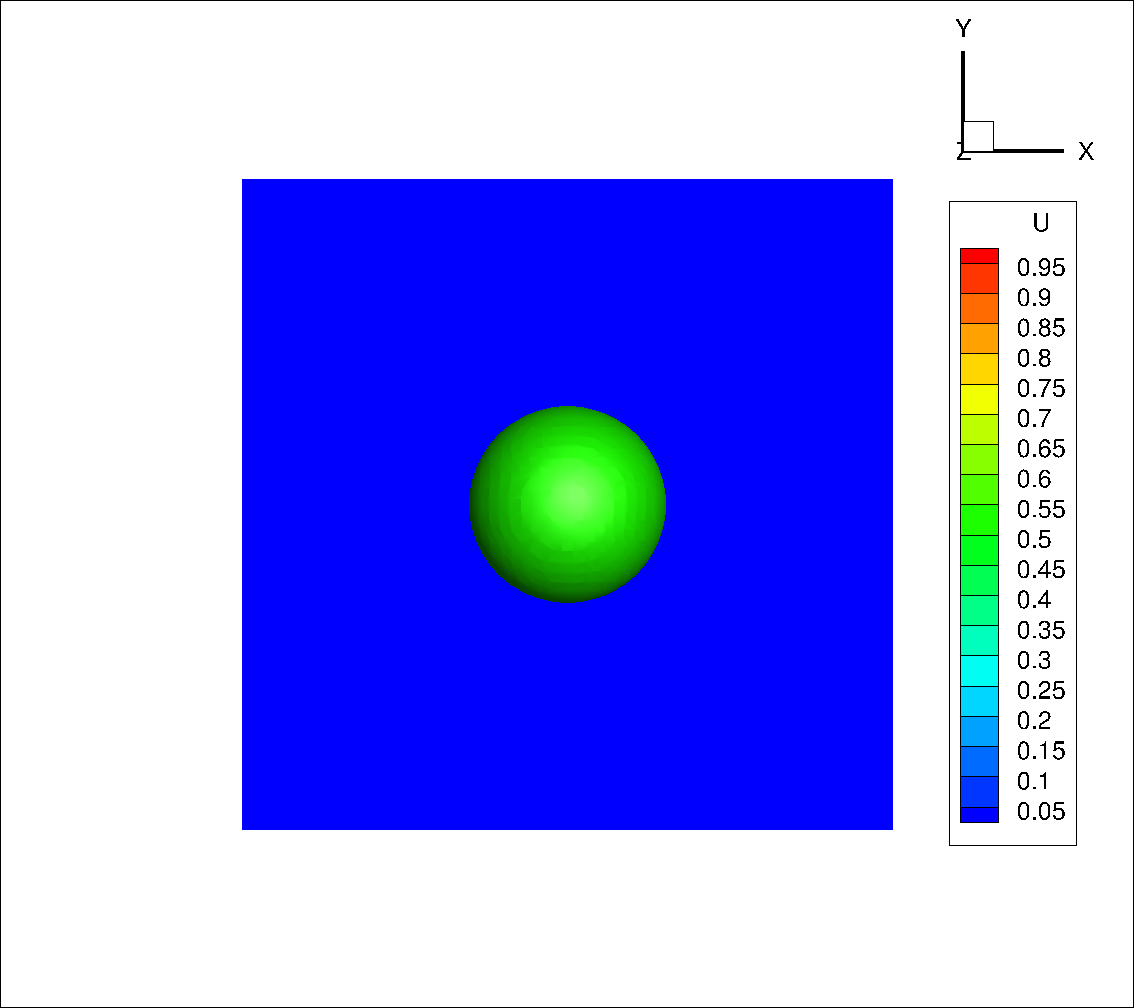} 
   \caption{}
  \label{fig:final_inconsist_drop_advection_mom_cons}  
   \end{subfigure}
    \caption{Velocity components at $t=1$ shown on top of the drop profile for drop advection with uniform velocity of $(0,0,1)$ in the domain (Case I). Shown is (a) $w$ from Equation \ref{mom_conservative_inconsistent}, (b) $w$ from Equation \ref{mom_con}, (c) $u$ from Equation \ref{mom_conservative_inconsistent}and (d) $u$ from Equation \ref{mom_con}.}
       \label{fig:drop_advection_mom_cons_vs_incons_vel}
\end{figure}

In this test case, we demonstrate the failure of merely using the conservative form of momentum equation (Equation \ref{momentum_generic_one_fluid_conservative}) with no accounting for the artificial mass fluxes. Namely, if instead of Equation \ref{mom_con}, we spatially and temporally discretize
\begin{equation}
    \frac { \partial (\rho \mathbf { u } ) }{ \partial t } +\nabla \cdot (\rho \mathbf { u } \otimes \mathbf { u } )=-\nabla P+\nabla \cdot \left[\mu (\nabla \mathbf { u } +\nabla ^{ T }\mathbf { u } ) \right ]+{\vec{ F }}_{ ST },
    \label{mom_conservative_inconsistent}
\end{equation}
which we refer to as the inconsistent method, the inconsistencies between mass and momentum fluxes result in huge errors. To demonstrate this, we consider simple simulations of drop
advection performed by discretely solving Equations \ref{mom_conservative_inconsistent} and \ref{mom_con} for momentum transport using central differences and RK4 time-stepping. A $D=0.3$ drop of fluid 1 is advected in a
periodic $1\times1\times1$ domain otherwise filled with fluid 2. The density of the fluids are chosen to be ${\rho}_{1}=10$ and ${\rho}_{2}=1$, while viscosity of both phases and surface tension is zero. The initial velocity is $(u,v,w)=(0,0,1)$ everywhere in the domain, and the mesh is $64\times64\times64$. Analytically, the drop should remain spherical during advection, and the velocity vector should not change. In Figure \ref{fig:drop_advection_mom_cons_vs_incons}, the initial profile of the drop along with the result of advecting the drop for one period (to $t=1$) using Equations \ref{mom_conservative_inconsistent} and \ref{mom_con} is shown. Clearly, while the drop is kept almost spherical with Equation \ref{mom_con}, the profile of the drop advected with the momentum-conserving but inconsistent momentum transport model given by Equation \ref{mom_conservative_inconsistent} is distorted and incorrect.
To further analyze this problem, we can look at the velocity field to understand why the drop profile is distorted when Equation \ref{mom_conservative_inconsistent} is solved. In Figure \ref{fig:drop_advection_mom_cons_vs_incons_vel} the value of velocity in the $x$ and $z$ direction are displayed at $t=1$ on top of the drop profile. Recall that the initial velocity was given by $(u,v,w)=(0,0,1)$, and the final velocity field using Equation \ref{mom_con} seems to have preserved that velocity field. On the other hand, the inconsistent momentum transport equation has altered the velocity field significantly. This is precisely what is meant by an inconsistent momentum advection scheme. It must be noted that had we chosen ${\rho}_{2}$ to be equal to ${\rho}_{1}$, both methods would have delivered correct solutions for this test. 

Applying the non-conservative form of the momentum transport equation (Equation \ref{NS}) to the above test case also results in $(u,v,w)(t)=(0,0,1)$ and a drop profile similar to the results obtained from Equation \ref{mom_con}. This would be the case for even extremely high density ratios. However, in what follows, we add one level of complexity to the test case, to see that Equation \ref{NS} also fails to produce acceptable results at high density ratios.

\subsection{Case II}
We consider a test case that is inspired by its 2D version presented in \cite{Bussmann2002} and \cite{Raessi2012}. In a periodic $1\times1\times1$ box of initially stationary fluid 2, a dense drop of fluid 1 with diameter $D=0.2$ is advected in the $x$-direction with initial velocity $(u,v,w)=(1,0,0)$. The spatial resolution is chosen to be $64\times64\times64$, and no surface tension or viscous forces are present. We consider very high density ratios up to  $\rho_{1}/\rho_{2}={10}^{7}$ to test the robustness of the solvers. There is no exact boundary for the drop in DI simulations, and the density is very high within the transition zone. In this case, the velocity in the domain is initialized using a hyperbolic tangent profile where the $U=0.5$ contour lies on points where the density is not very high, say ${10}^{3}{\rho}_{2}$. A CFL number of 0.25 is chosen for these simulations. Theoretically, when the density ratio is very high, the drop should not "feel" the presence of the surrounding phase and should not deform. The nonconservative formulation of Equation \ref{NS} and our modified momentum transport equation given in Equation \ref{mom_con} are compared for this test case.

\begin{figure} 
\centering
\begin{subfigure}{0.38 \linewidth}
\includegraphics[width=0.95\textwidth]{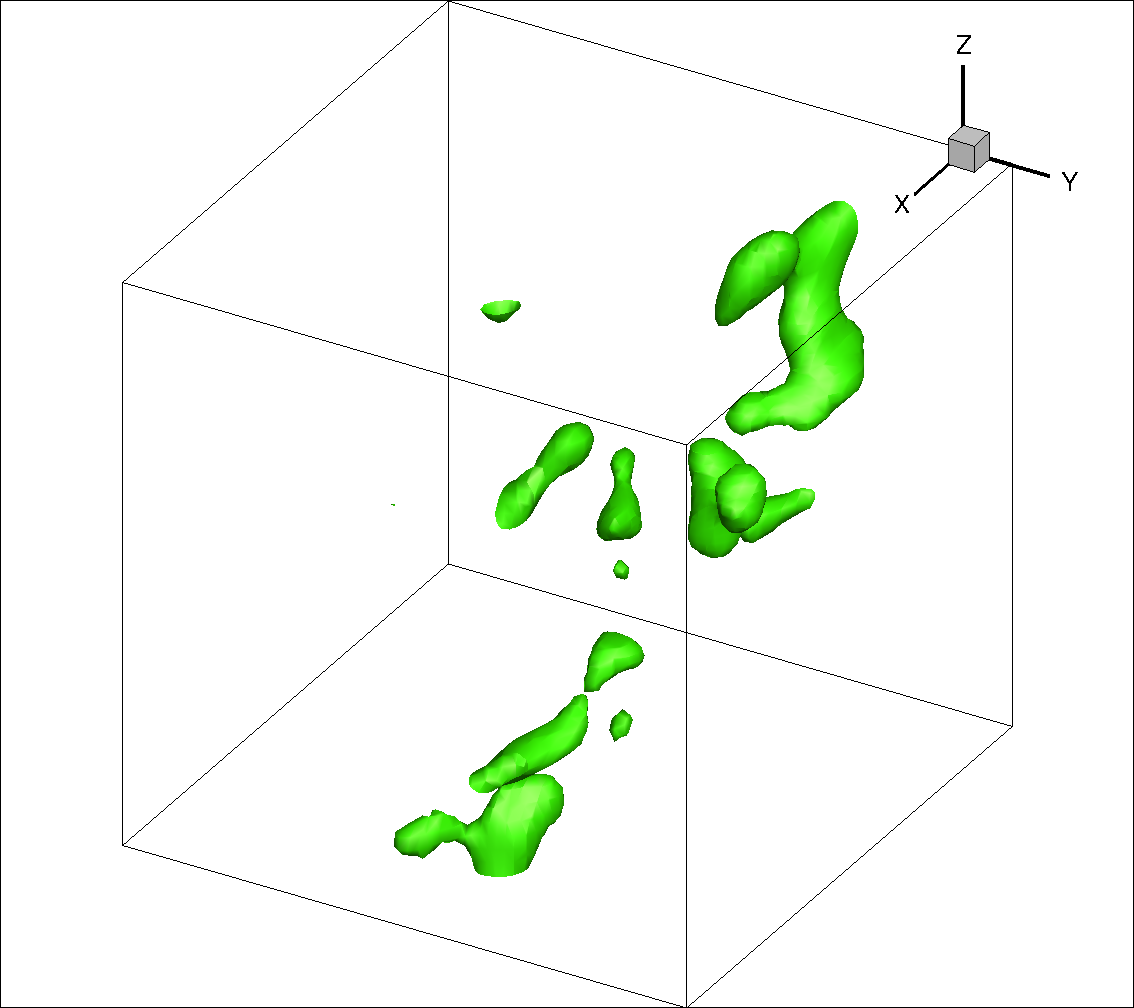} 
   \caption{}
   \label{fig:init_drop_advection_mom_cons} 
\end{subfigure}
\begin{subfigure}{0.38 \linewidth}
\includegraphics[width=0.95\textwidth]
{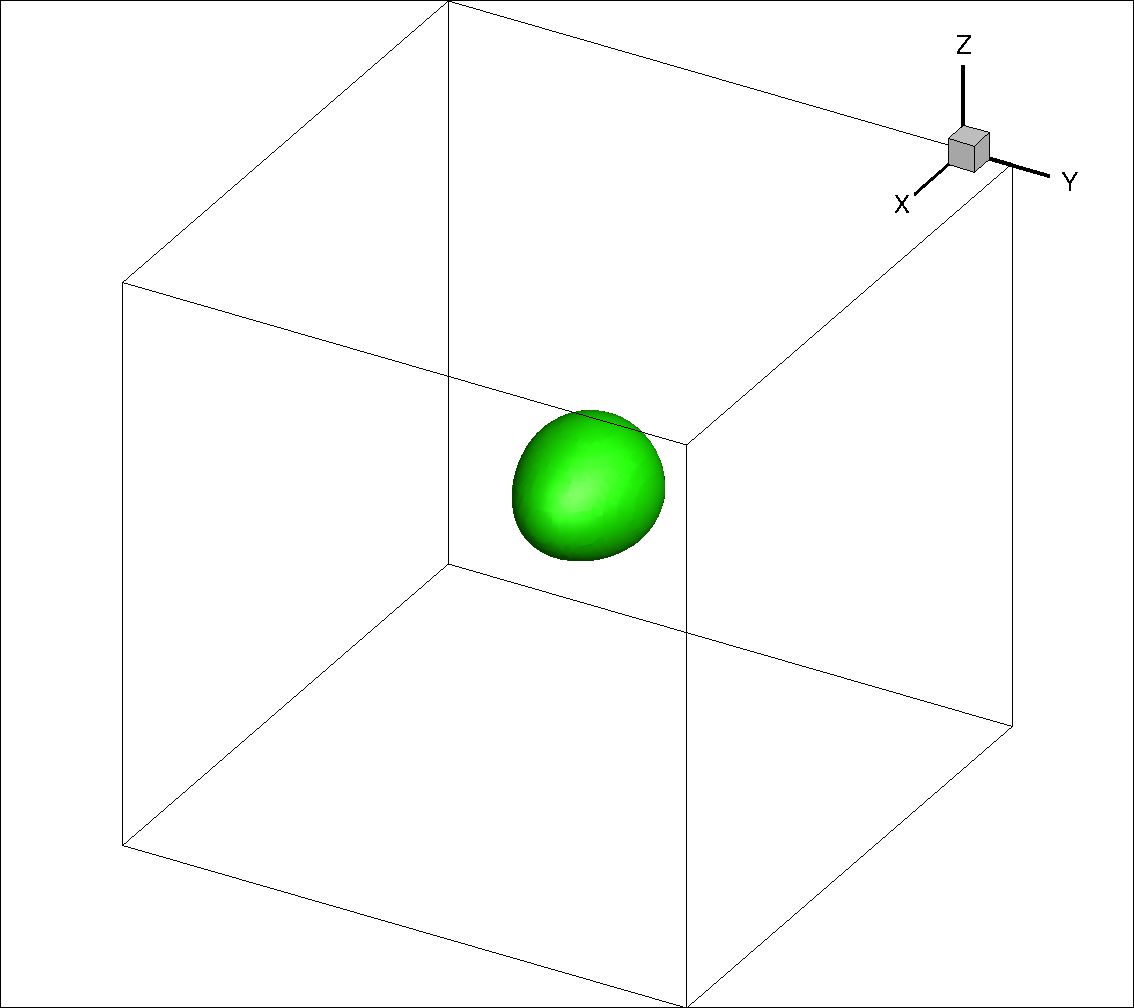} 
   \caption{}
  \label{fig:final_inconsist_drop_advection_mom_cons}  
   \end{subfigure}
   \begin{subfigure}{0.38 \linewidth}
\includegraphics[width=0.95\textwidth]
{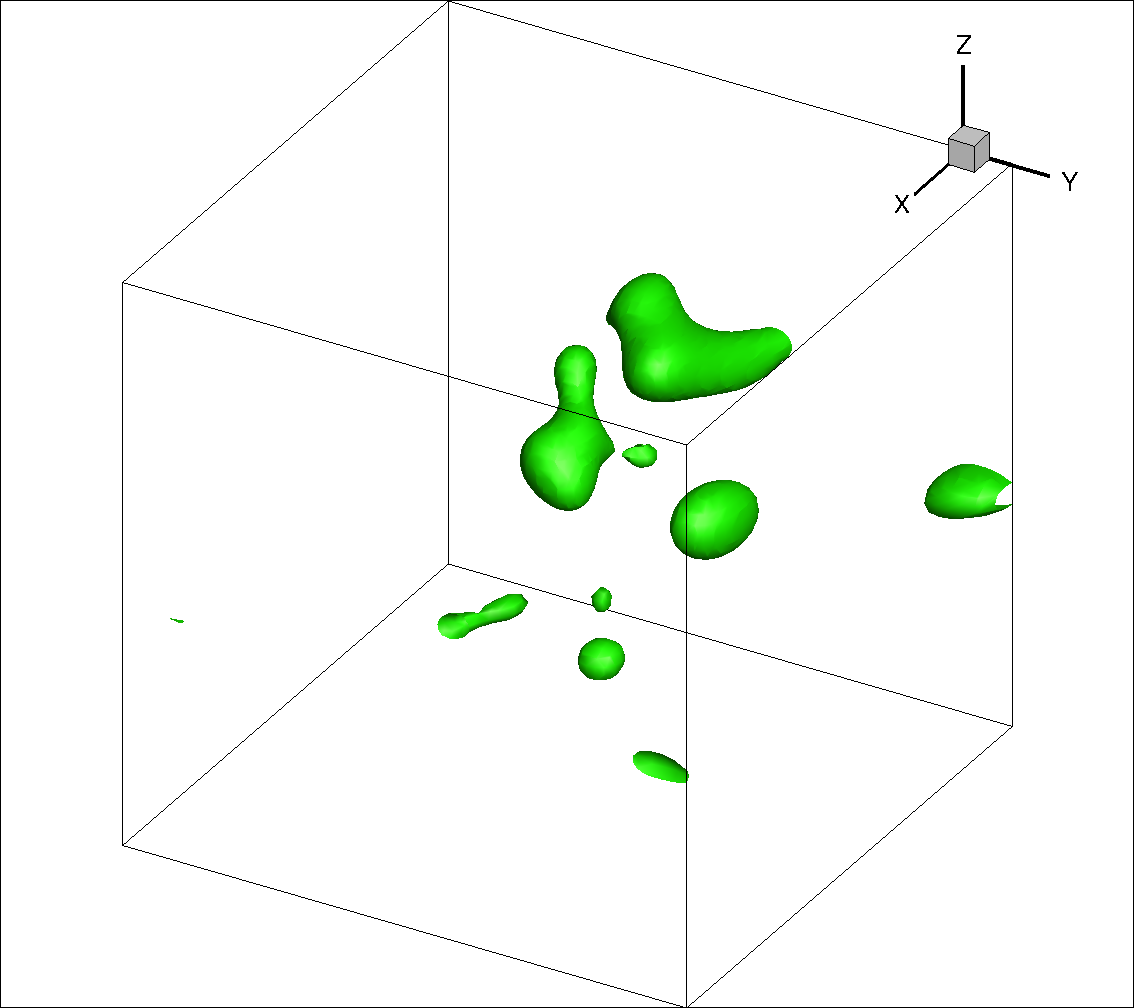} 
   \caption{}
  \label{fig:final_inconsist_drop_advection_mom_cons}  
   \end{subfigure}
      \begin{subfigure}{0.38 \linewidth}
\includegraphics[width=0.95\textwidth]
{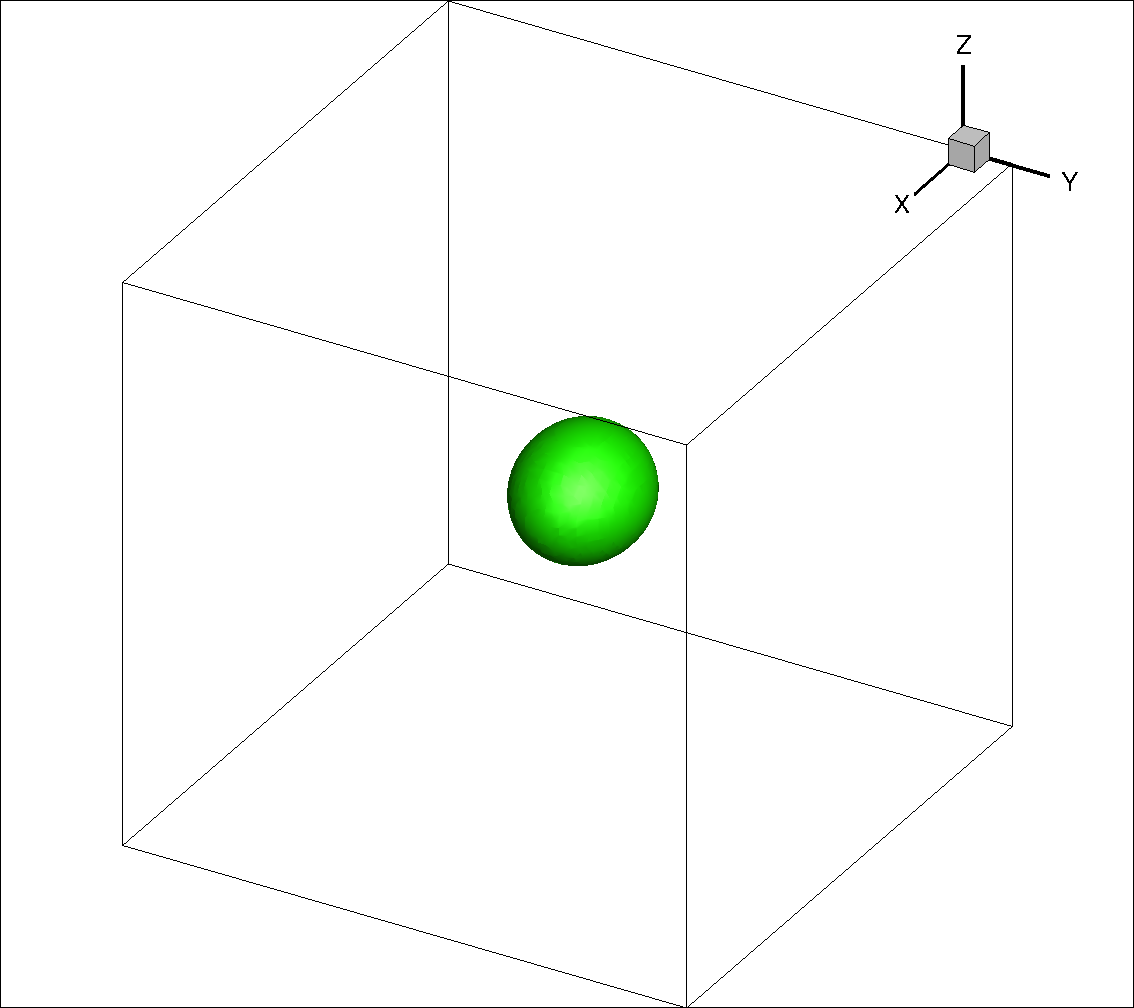} 
   \caption{}
  \label{fig:final_inconsist_drop_advection_mom_cons}  
   \end{subfigure}
      \begin{subfigure}{0.38 \linewidth}
\includegraphics[width=0.95\textwidth]
{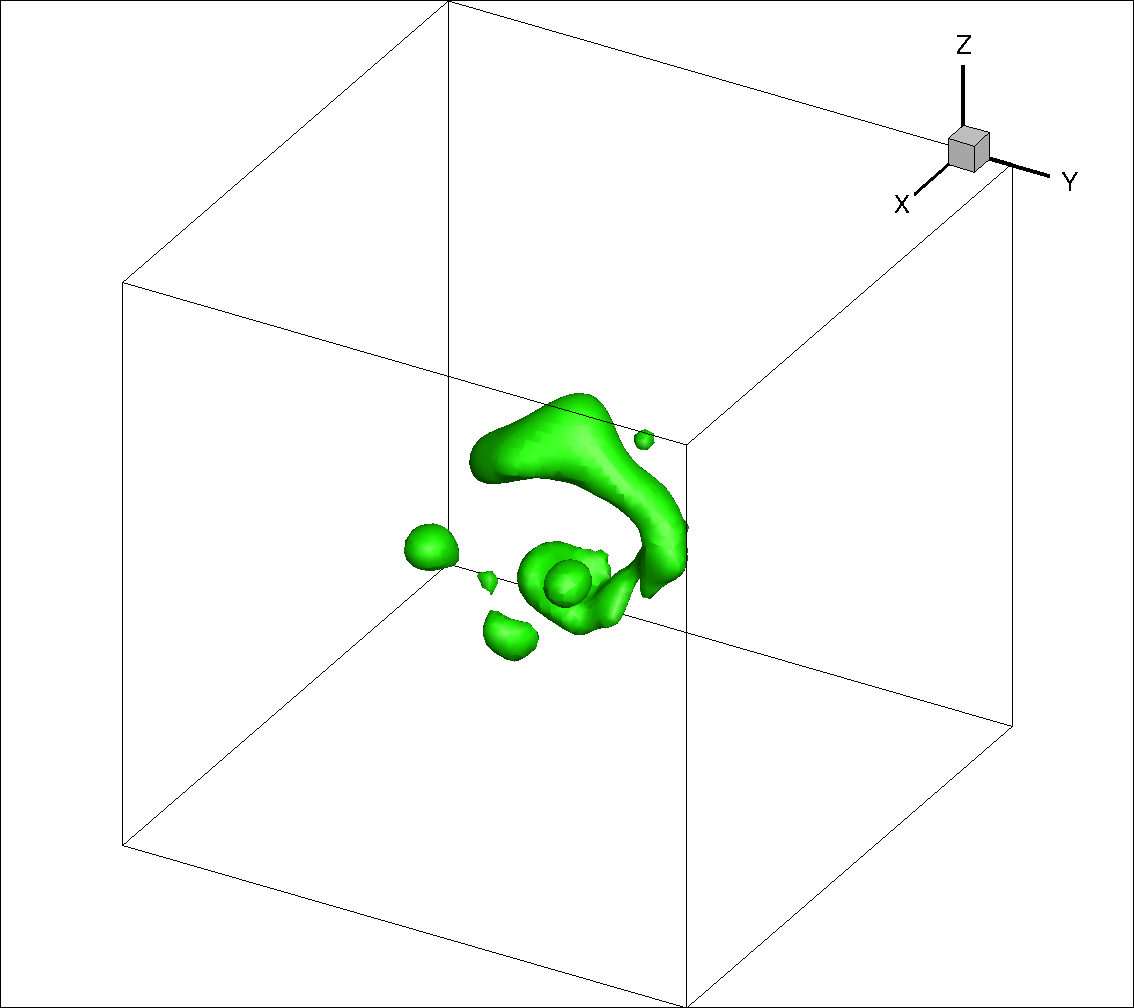} 
   \caption{}
  \label{fig:final_inconsist_drop_advection_mom_cons}  
   \end{subfigure}
      \begin{subfigure}{0.38 \linewidth}
\includegraphics[width=0.95\textwidth]
{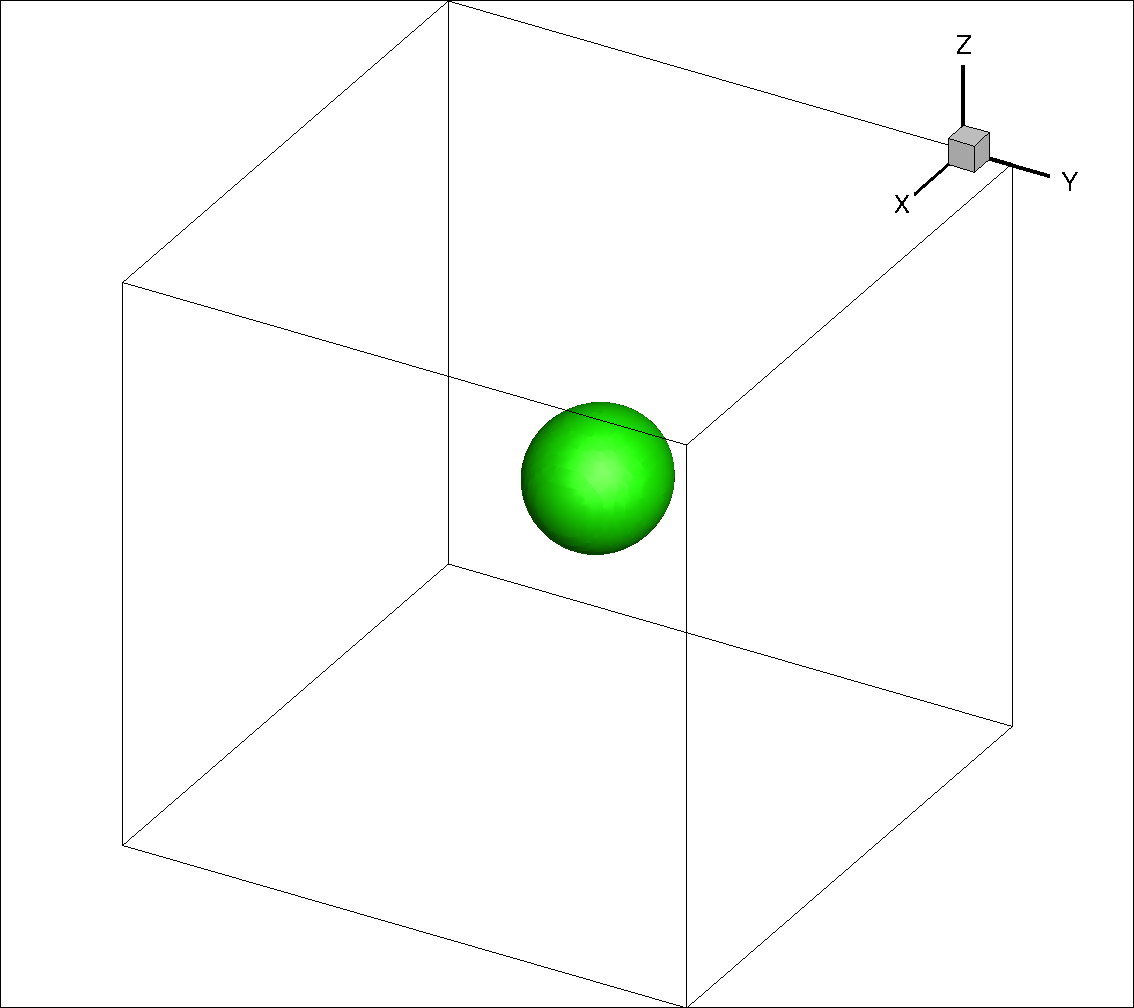} 
   \caption{}
  \label{fig:final_inconsist_drop_advection_mom_cons}  
   \end{subfigure}
    \caption{Drop profile shown for case of a dense drop advected in an initially stationary background (Case II) for (a) density ratio of ${10}^{4}$ at $t=8.41$ with Equation \ref{NS}, (b) density ratio of ${10}^{4}$ at $t=10$ with Equation \ref{mom_con}, (c) density ratio of ${10}^{6}$ at $t=2.72$ with Equation \ref{NS}, (d) density ratio of ${10}^{6}$ at $t=10$ with Equation \ref{mom_con}, (e) density ratio of ${10}^{7}$ at $t=2.01$ with Equation \ref{NS}, (f) density ratio of ${10}^{7}$ at $t=10$ with Equation \ref{mom_con}.}
       \label{fig:dense_drop}
\end{figure}

By means of these simulations, we observe that the simulations performed using Equation \ref{NS} are not robust at density ratios of ${10}^{4}$ and above, whereas simulations with Equation \ref{mom_con} are stable even at a density ratio of ${10}^{7}$ and successfully capture the theoretical solution. The drop profiles at various high density ratios are shown in Figure \ref{fig:dense_drop}. The time of the snapshots for the simulations with our conservative and consistent method are all at $t=10$, while for the non-conservative momentum solver, the time of the snapshot is shortly before the solver becomes numerically unstable (at $t<10$).
 
\begin{figure}
\centering
\includegraphics[width=0.75\textwidth]{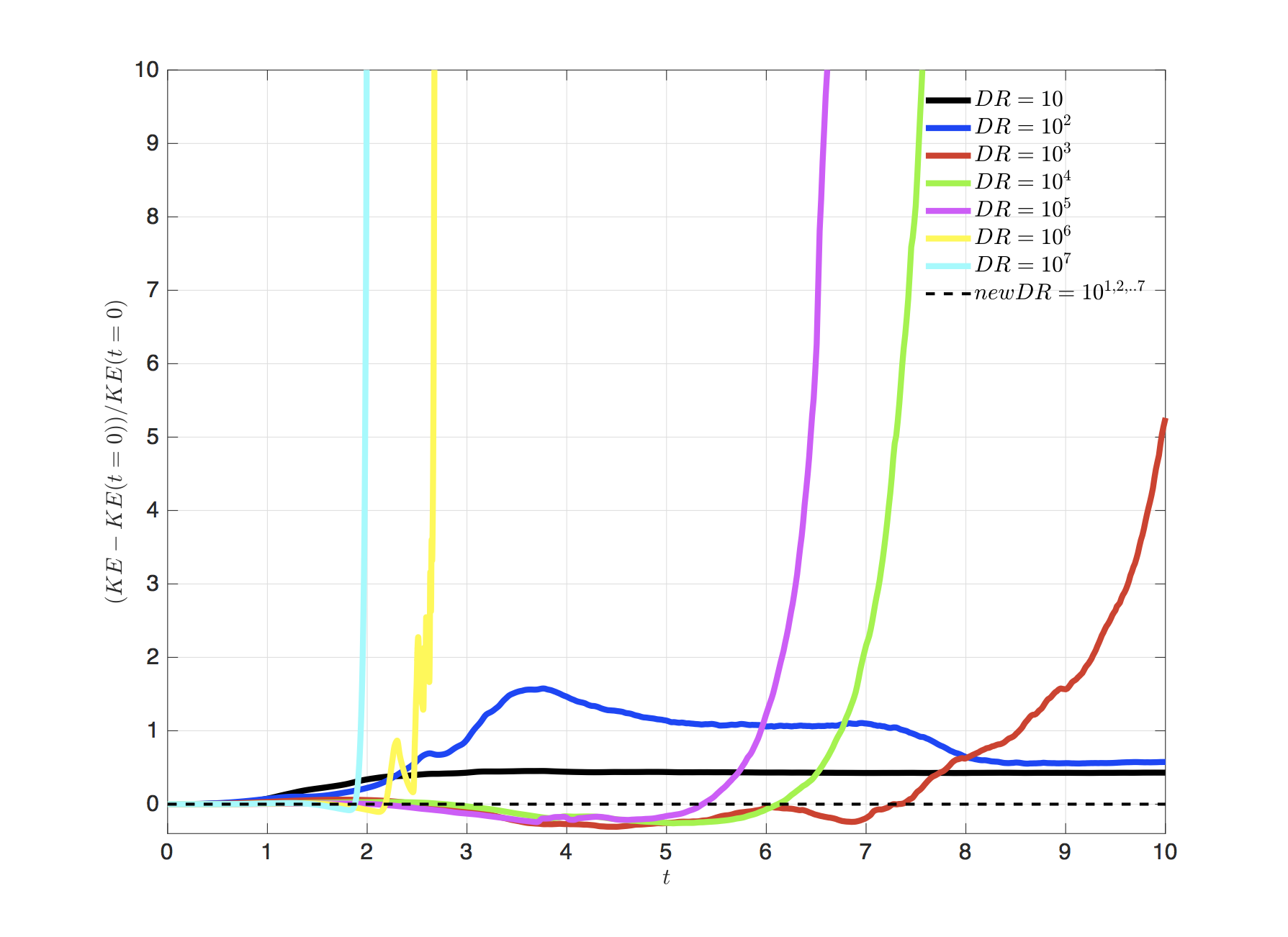}
\caption{KE conservation error of the system in the dense drop advection case plotted for different density ratios. Different solid lines represent simulations using Equation \ref{NS} for various density ratios, while the dashed line represents the very small (on the order of Poisson tolerance-${10}^{-12}$) KE conservation error for all density ratios using Equation \ref{mom_con}.}
\label{fig:KE_drops}
\end{figure}

We can also examine the evolution of discrete kinetic energy of the system, defined in Equation \ref{KE_discrete_defn}, in these simulations. In Figure \ref{fig:KE_drops}, we demonstrate how radically the kinetic energy of the system changes when simulating even low density ratio simulations of this test case with Equation \ref{NS}. Crucially, we can see how Equation \ref{NS} is numerically unstable at high density ratios. On the other hand, it is clear that the total kinetic energy of the system is constant for simulations that solve Equation \ref{mom_con} for momentum transport. This is expected, as in the absence of surface tension and viscous forces, we have proven in Section \ref{sec:proof_KE_conservation} that Equation \ref{mom_con} conserves total kinetic energy. Errors in kinetic energy conservation are small and due merely to the fact that the continuity equation (Equation \ref{incompressible_semidiscrete} in 2D) is enforced with a non-zero tolerance when numerically solving the Poisson pressure system.

It must be noted that increasing mesh resolution while solving Equation \ref{NS} coupled to Equation \ref{phitrans} does not improve conservation errors or robustness of the solver. We have examined this for the dense drop advection problem, in addition to the jet in cross-flow simulations presented in the next section.

\subsection{Jet in cross-flow simulation} 
\label{sec:jet_crossflow}

While there is no shortage of new studies on phase field methods, we rarely see these methods being applied to realistic engineering problems. In most articles in fact, the authors study canonical problems or flows with low $Re$ numbers or not so high density ratios. In this section, we employ our diffuse interface method to simulate the realistic problem of a jet in cross-flow. This is a suitable problem to examine the practical importance of using Equation \ref{mom_con} for momentum transport. To this aim, we have performed simulations at various density ratios to compare the solutions obtained from solving Equation \ref{mom_con} or \ref{NS} for momentum transport. Both of these equations are coupled to Equation \ref{phitrans} and are solved in the same framework.

The inlet conditions are laminar for the jet but the flow becomes turbulent downstream. Experimental measurements for this problem using water jets in atmospheric conditions have been performed by \cite{Sallam2004} among others. This problem has also been studied numerically using a CLSVOF scheme in \cite{Li2012}. It should be noted that primary breakup in this problem is due to the gas cross-flow and not turbulence or vorticity in the jet. Furthermore, the specific case selected for simulation belongs to the multi-mode breakup regime which has the most interesting dynamics, has relevance to aerospace applications, and requires reasonable mesh resolution. 

In the simulations presented in \cite{Li2012}, the authors used a uniform grid or adaptive mesh refinement to construct the mesh. In our study, we use a non-uniform staggered grid which is concentrated and has uniform spacing around the jet. In Section \ref{sec:non_uniform_mesh}, we will explain how Equations \ref{phitrans}, \ref{mom_con} and \ref{NS} must be discretized for a non-uniform Cartesian mesh. In particular, we will explain how we are able to achieve the following important properties on non-uniform meshes:
\begin{itemize}
    \item Equation \ref{phitrans} is discretized in a manner that the interface thickness varies in space according to the local mesh, while staying within the $0-1$ bounds. This is a superior option compared to choosing $\epsilon$ based on the coarsest mesh in the domain, especially in scenarios where the momentum transport equation must be solved in a domain much larger than where the interfaces reside (e.g. the jet in cross-flow case).
    \item Similar to \cite{Ham2002} who showed how one can achieve a fully conservative second order finite difference scheme for incompressible single phase flows on non-uniform grids, by using volume-weighted interpolation on the convective terms, we can extend Equation \ref{KE_discrete_conservation} to non-uniform grids.
\end{itemize}

The geometric specification of the jet and domain, in addition to boundary conditions, are chosen following \cite{Li2012}. The domain size is $2 cm\times1.5 cm\times3.5 cm$, where the cross-flow blows in the positive $x$ direction and the jet is injected in the $y$ direction. The jet orifice has a diameter of ${D}_{ori}=0.8 mm$ and is situated at $(0.2 cm, 0.75 cm, 0 cm)$. Inflow boundary conditions are used for the gas inflow at $x=0 cm$, in addition to the jet inflow at the jet orifice. For all other points at $y=0 cm$, no-slip boundary conditions are imposed. For the downstream boundary at $x=2 cm$, convective outflow boundary conditions are implemented, while no penetration, free-slip boundary conditions are used on all the other boundaries of the domain. The domain is sufficiently large in the $y$ and $z$ direction such that the boundary conditions in the $z$ direction and the top wall do not influence the results of our simulations. The initial conditions for velocity are $(u,v,w)(\vec{x})=({U}_{g},0,0)$, for all $x$ inside the domain (not on boundaries). The liquid jet is thus injected into this uniform velocity cross-flow at $t=0$. 

The dimensional and non-dimensional parameters governing this problem can be seen in Table \ref{tab:jet_params}. In this table, in addition to the familiar parameters, the momentum flux ratio, $q={\rho}_{l}{U}_{l}^{2}/({\rho}_{g}{U}_{g}^{2})$ is used to characterize the problem. We have explored two cases, with different density ratios but matched ${Re}_{l}$, ${We}_{g}$, $q$ and identical geometric configurations. For this to be possible, the viscosity ratio between the two cases must be different as well. Table \ref{tab:jet_params} fully specifies these two cases. All dimensional parameters are reported in metric units. It must be noted that both of these density ratios are physically relevant, as even the reduced density of ratio of $100$ is common for a liquid jet that is injected into a pressurized gas chamber. 

\begin{table}[H]
\begin{center} 
\scalebox{0.8}{
\begin{tabular}{c|c|c|c|c|c|c|c|c|c|c|c|c|c}
Case  & ${\rho}_{l}$ & ${\rho}_{g}$ & ${\mu}_{l}$ & ${\mu}_{g}$ & $\sigma$ & ${D}_{ori}$ & ${U}_{l}$ & ${U}_{g}$ & ${\rho}_{l}/{\rho}_{g}$ & ${\mu}_{l}/{\mu}_{g}$ & ${We}_{g}$ & ${Re}_{l}$ & $q$\\ \hline
1 & 118 & 1.18 & 0.000307 & 0.0000186 & 0.0708 & 0.0008 & 51.45 & 54.8 & 100 & 16.51 & 40 & 15800 & 88.2 \\
2 & 997 & 1.18 & 0.000894 & 0.0000186 & 0.0708 & 0.0008 & 17.7 & 54.8 & 845 & 48 & 40 & 15800 & 88.2 \\
\end{tabular}}
\end{center}
\vspace{1mm}
\caption{Physical parameters and material properties for jet in cross-flow simulations.}
\label{tab:jet_params}
\end{table}

A summary of all the simulations that we examine herein is reported in Table \ref{tab:jet_sims}. In what follows, we go through these simulations and focus on comparing the two options for the momentum transport equation when applied to solving the same physical cases (Table \ref{tab:jet_params}).

\begin{figure}
\centering
\begin{subfigure}{0.49 \linewidth}
\includegraphics[width=\textwidth]{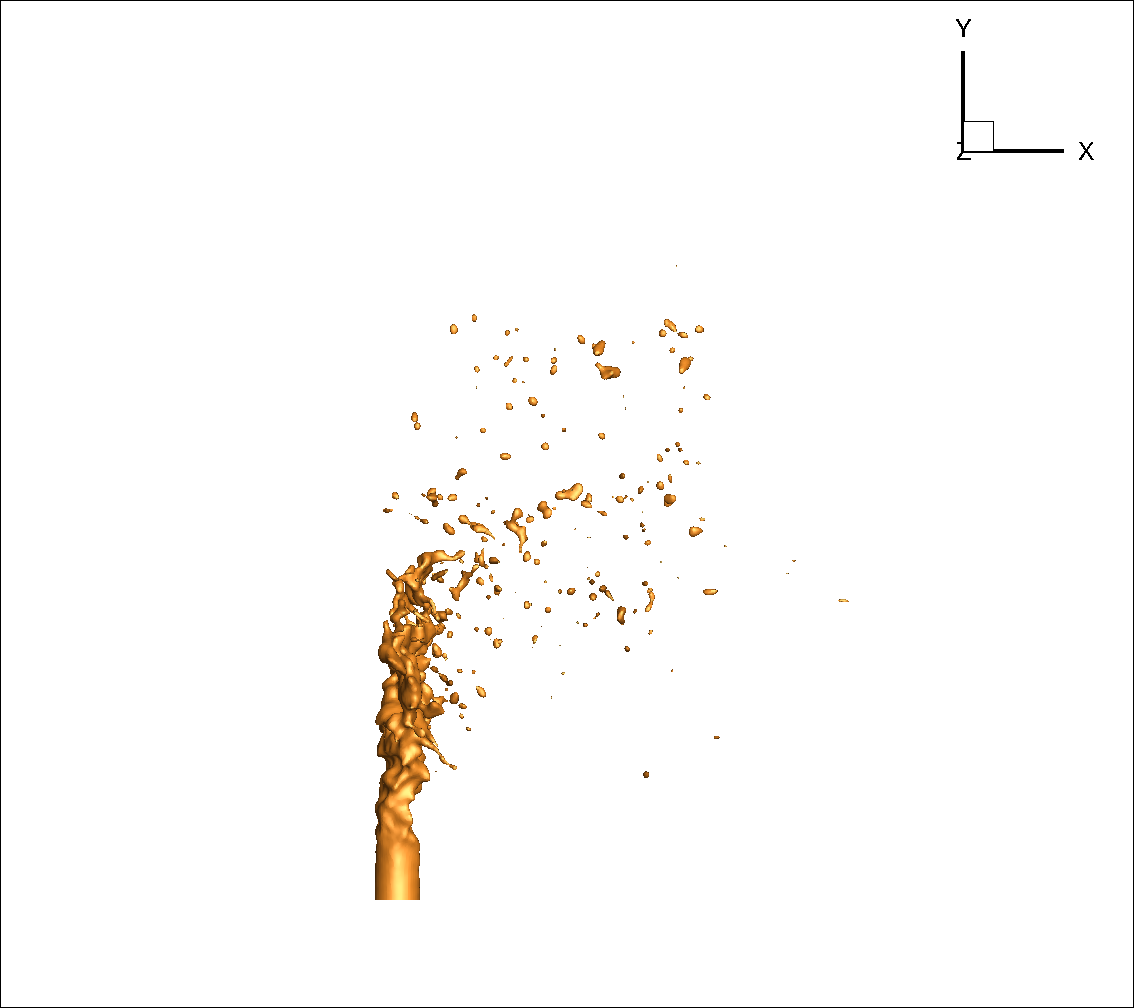} \caption{} 
 \label{fig:jet_non_xy}
\end{subfigure}
\begin{subfigure}{0.49 \linewidth}
\includegraphics[width=\textwidth]{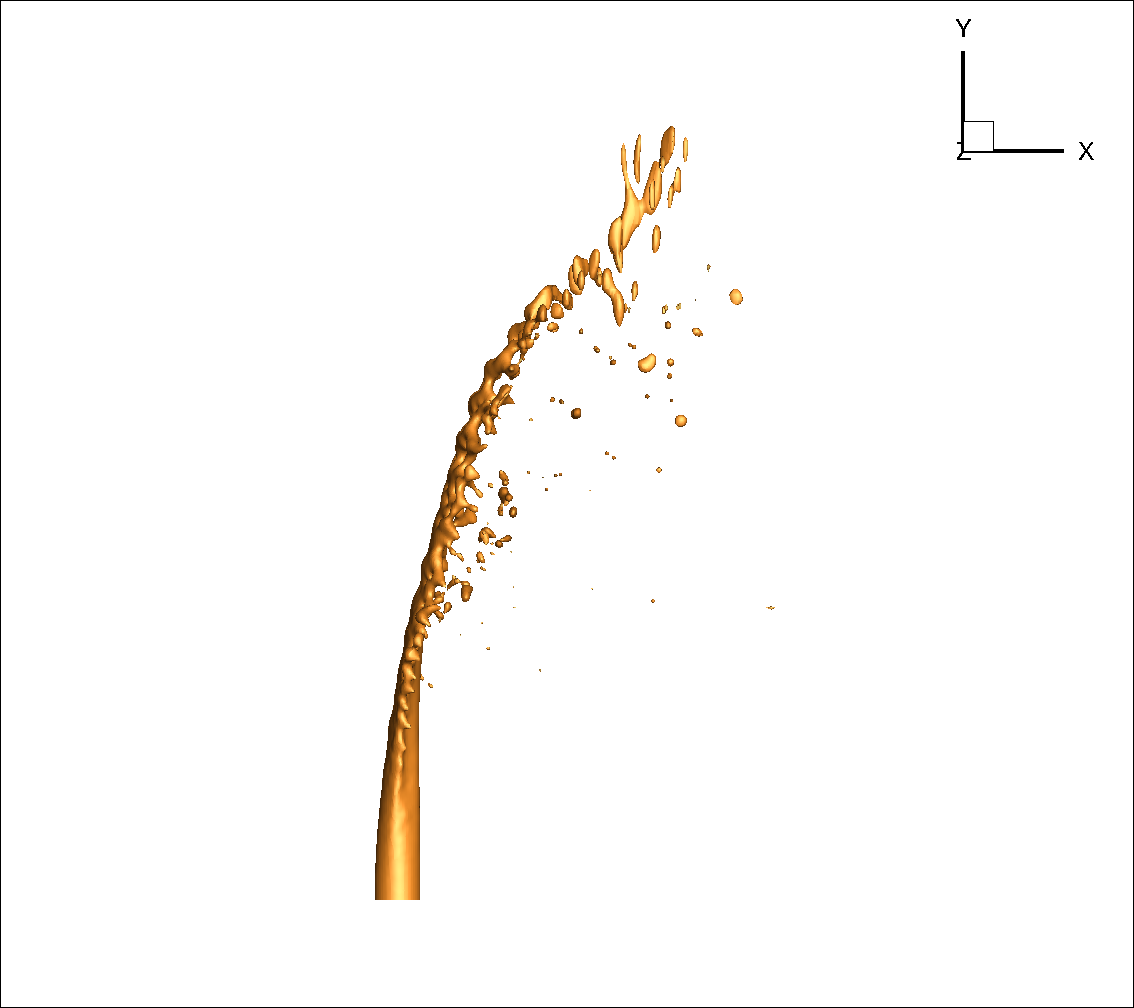} \caption{} 
 \label{fig:jet_con_xy} 
\end{subfigure}\\
\begin{subfigure}{0.49 \linewidth}
\includegraphics[width=\textwidth]{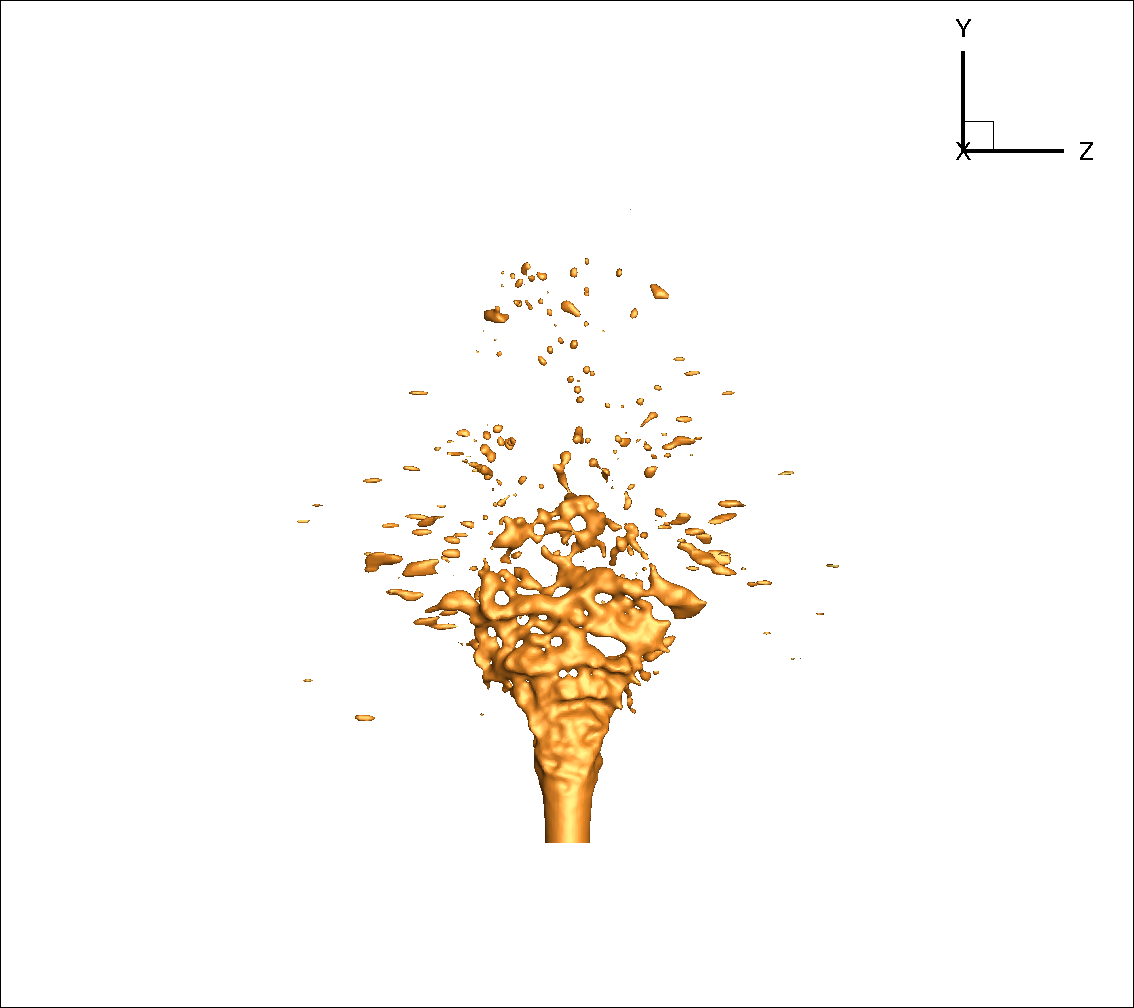} \caption{} 
 \label{fig:jet_non_yz}
\end{subfigure}
\begin{subfigure}{0.49 \linewidth}
\includegraphics[width=\textwidth]{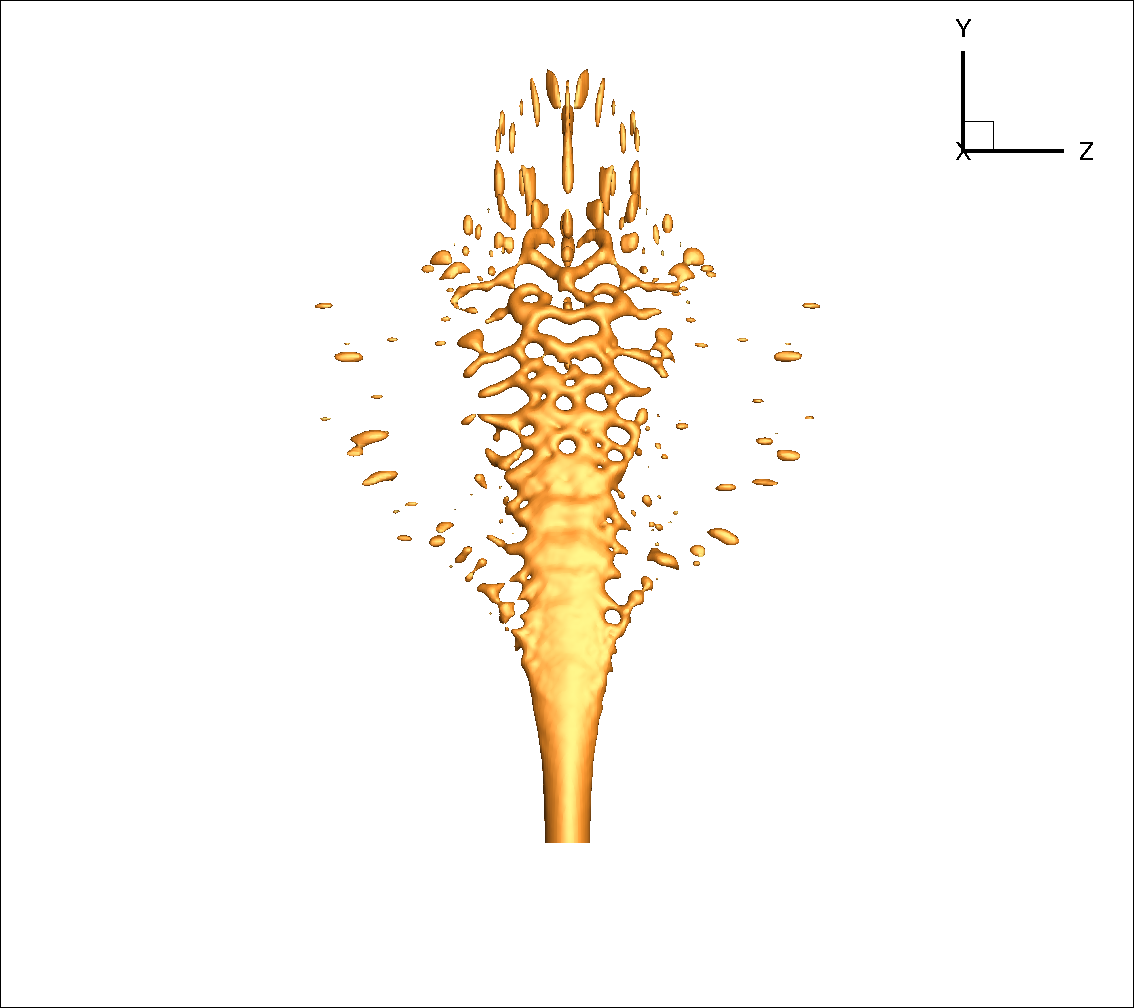} \caption{} 
 \label{fig:jet_con_yz} 
\end{subfigure}
\caption{Results of jet in cross-flow simulation of Case 1 using Equations \ref{NS} and \ref{mom_con} for momentum transport. Zoom-ins on the jet profile with (a) x-y view of Simuation 1 using Equation \ref{NS} (b) x-y view of Simuation 2 using Equation \ref{mom_con} (c) y-z view of Simuation 1 using Equation \ref{NS} (d) y-z view of Simuation 2 using Equation \ref{mom_con}.}
\label{fig:jet_case_1_coarse}
\end{figure}

\begin{table}[H]
\begin{center} 
\begin{tabular}{c|c|c|c|c}
Simulation & Case  & Mom. Eqn. & Resolution & Robustness\\ \hline
1 & 1 & nonconservative, Eq. \ref{NS} & ${D}_{ori}/24$ & stable\\
2 & 1 & conservative, Eq. \ref{mom_con} & ${D}_{ori}/24$ & stable\\
3 & 1 & nonconservative, Eq. \ref{NS} & ${D}_{ori}/36$ & unstable\\
4 & 1 & conservative, Eq. \ref{mom_con} & ${D}_{ori}/36$ & stable\\
5 & 2 & nonconservative, Eq. \ref{NS} & ${D}_{ori}/24$ & unstable\\
6 & 2 & conservative, Eq. \ref{mom_con} & ${D}_{ori}/24$ & stable\\
\end{tabular}
\end{center}
\vspace{1mm}
\caption{Summary of the set of simulations studied.}
\label{tab:jet_sims}
\end{table}
Simulations 1 and 2 in Table \ref{tab:jet_sims} are coarse simulations of Case 1, in which the density ratio is $100$. In Figure \ref{fig:jet_case_1_coarse}, a side-by-side comparison of the output of the simulations at $t=2.74\times{10}^{-4}s$ can be observed from two different views. Based on experimental images and simulation results at such $Re$, $q$ and $We$ in literature\citep{Sallam2004,Li2012}, it is clear that solving the conservative and consistent
version of the Navier-Stokes equation (Equation \ref{mom_con}) gives much more realistic break-up of the jet and jet column trajectory. The non-conservative form of Navier-Stokes (Equation \ref{NS}) leads to premature break-up of the jet and nonphysical perturbations early on in its trajectory. It should be reminded that both of these simulations are coarse and employ highly stretched meshes in order to have a uniform grid around the jet with 24 cells across its diameter. As such, artificially stretched drops can be observed especially in the case of simulation 2 where the jet
has reached higher into the stretched region of the mesh. These artifacts can be reduced via mesh refinement, using a uniform mesh, or a different surface tension force implementation tailored for non-uniform grids.

\begin{figure}
\centering
\begin{subfigure}{0.32 \linewidth}
\includegraphics[width=\textwidth]{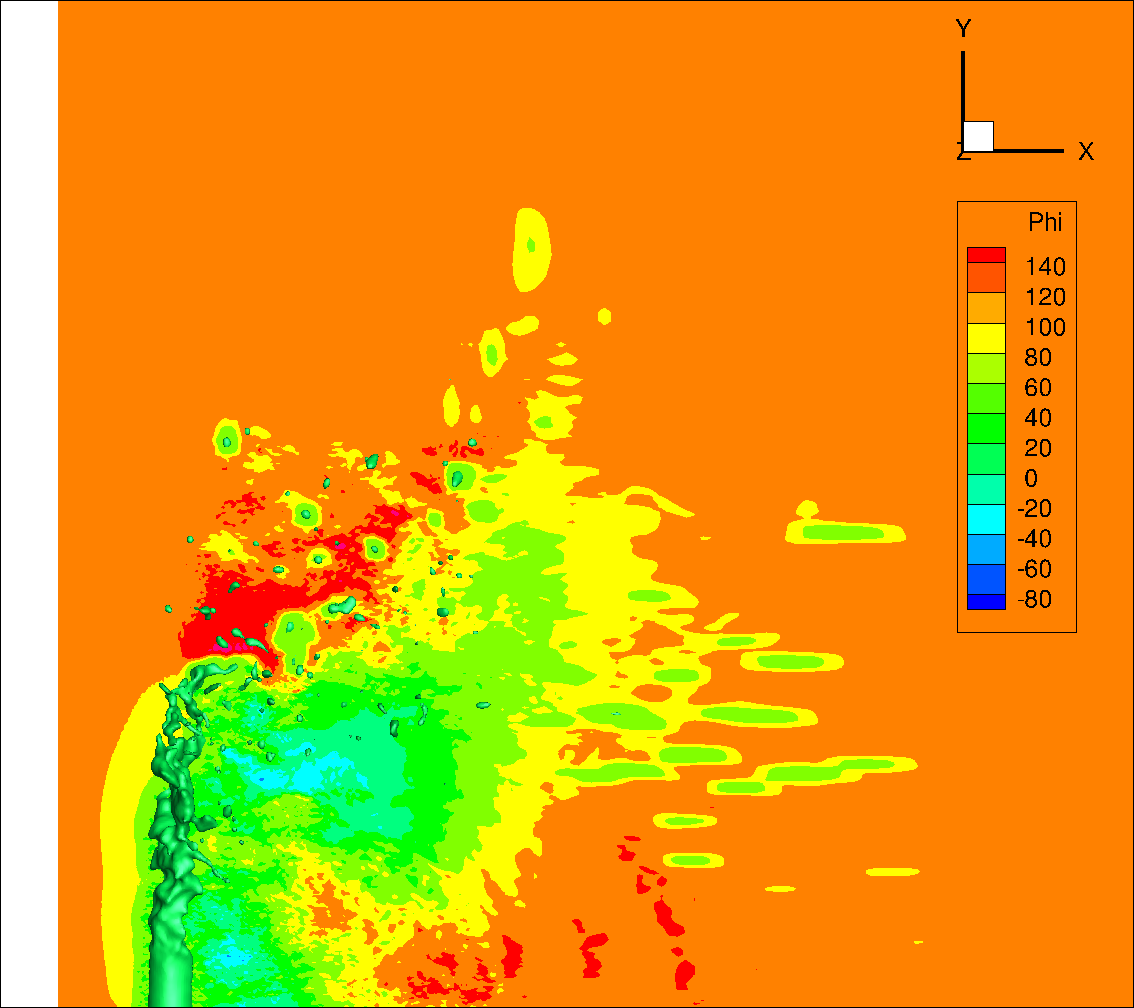} \caption{} 
 \label{fig:jet_non_xy}
\end{subfigure}
\begin{subfigure}{0.32 \linewidth}
\includegraphics[width=\textwidth]{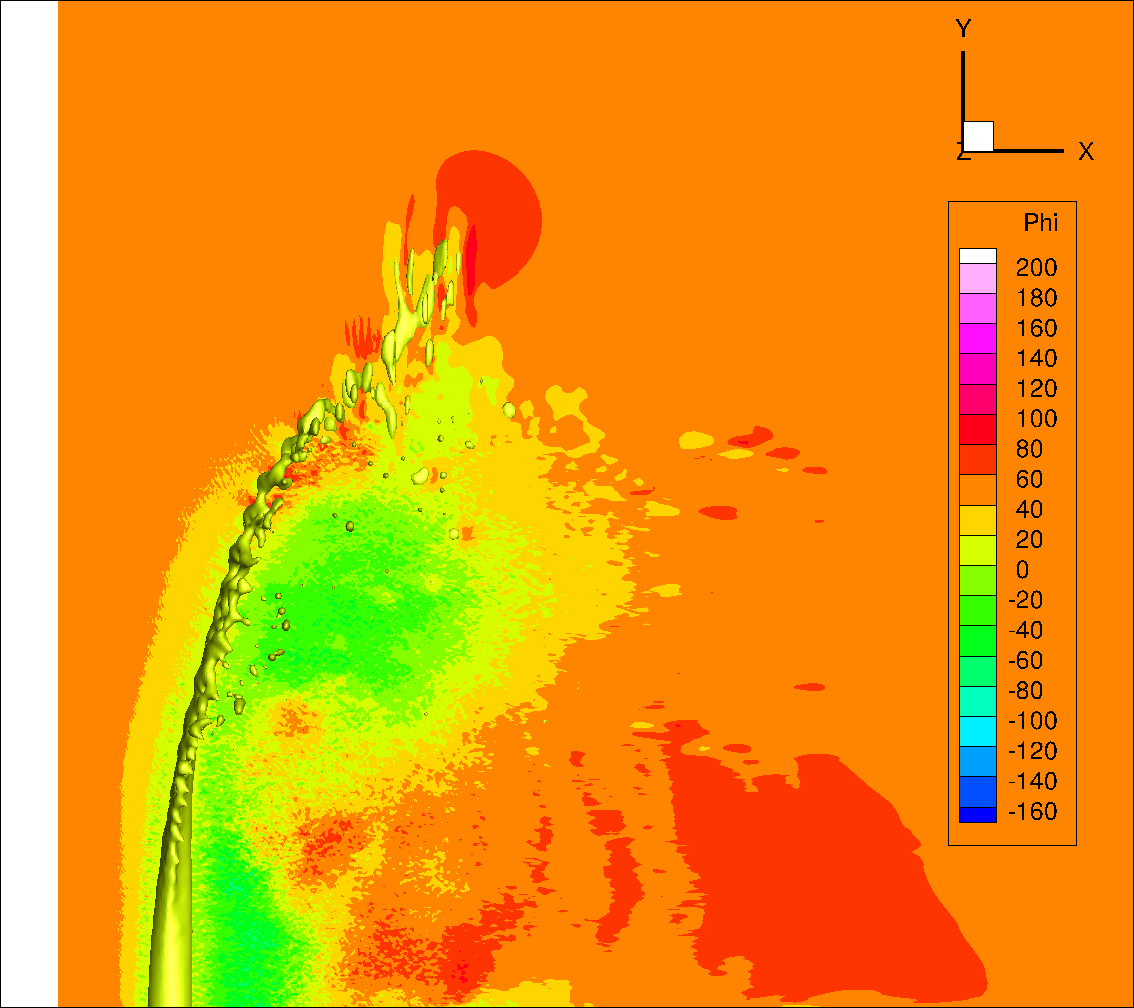} \caption{} 
 \label{fig:jet_con_xy} 
\end{subfigure}\\
\begin{subfigure}{0.32 \linewidth}
\includegraphics[width=\textwidth]{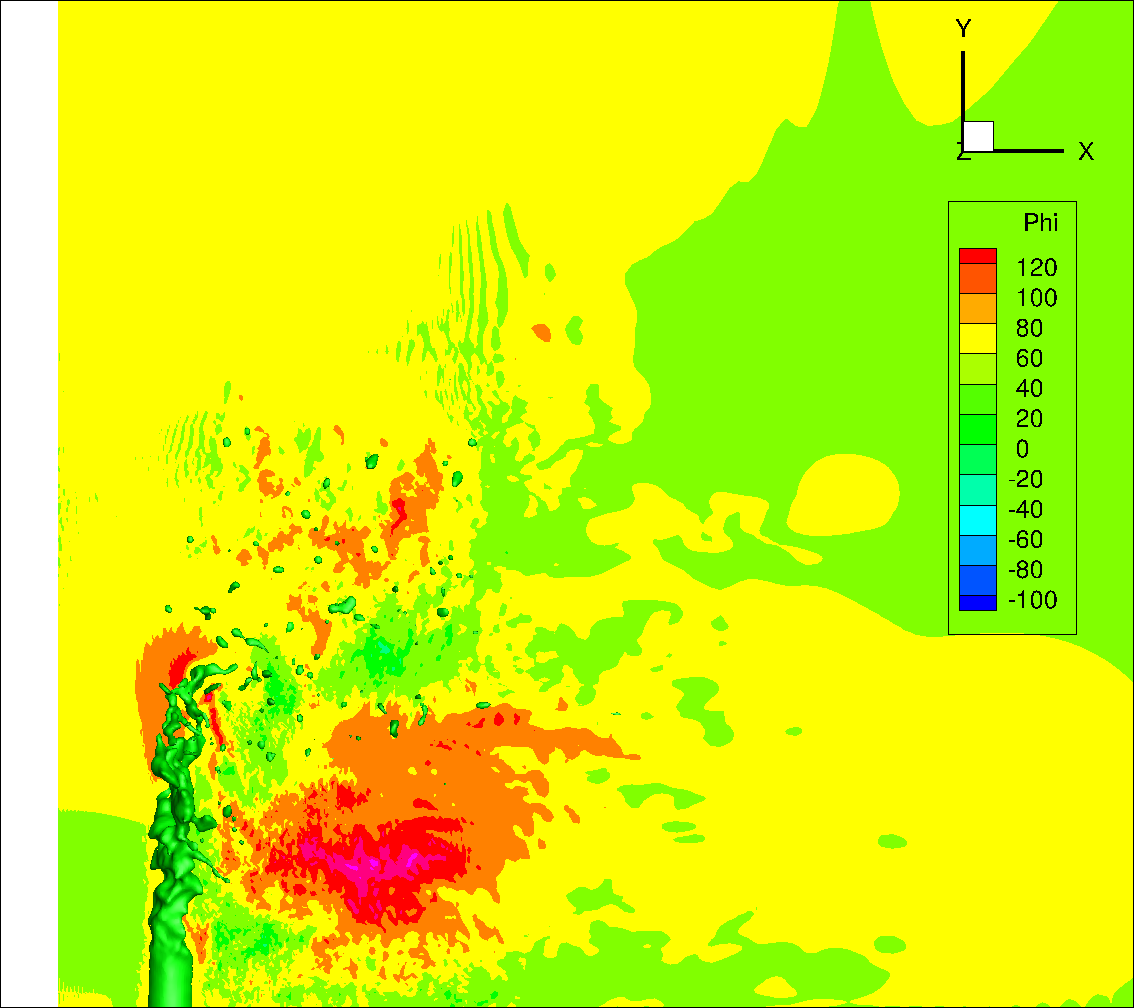} \caption{} 
 \label{fig:jet_non_yz}
\end{subfigure}
\begin{subfigure}{0.32 \linewidth}
\includegraphics[width=\textwidth]{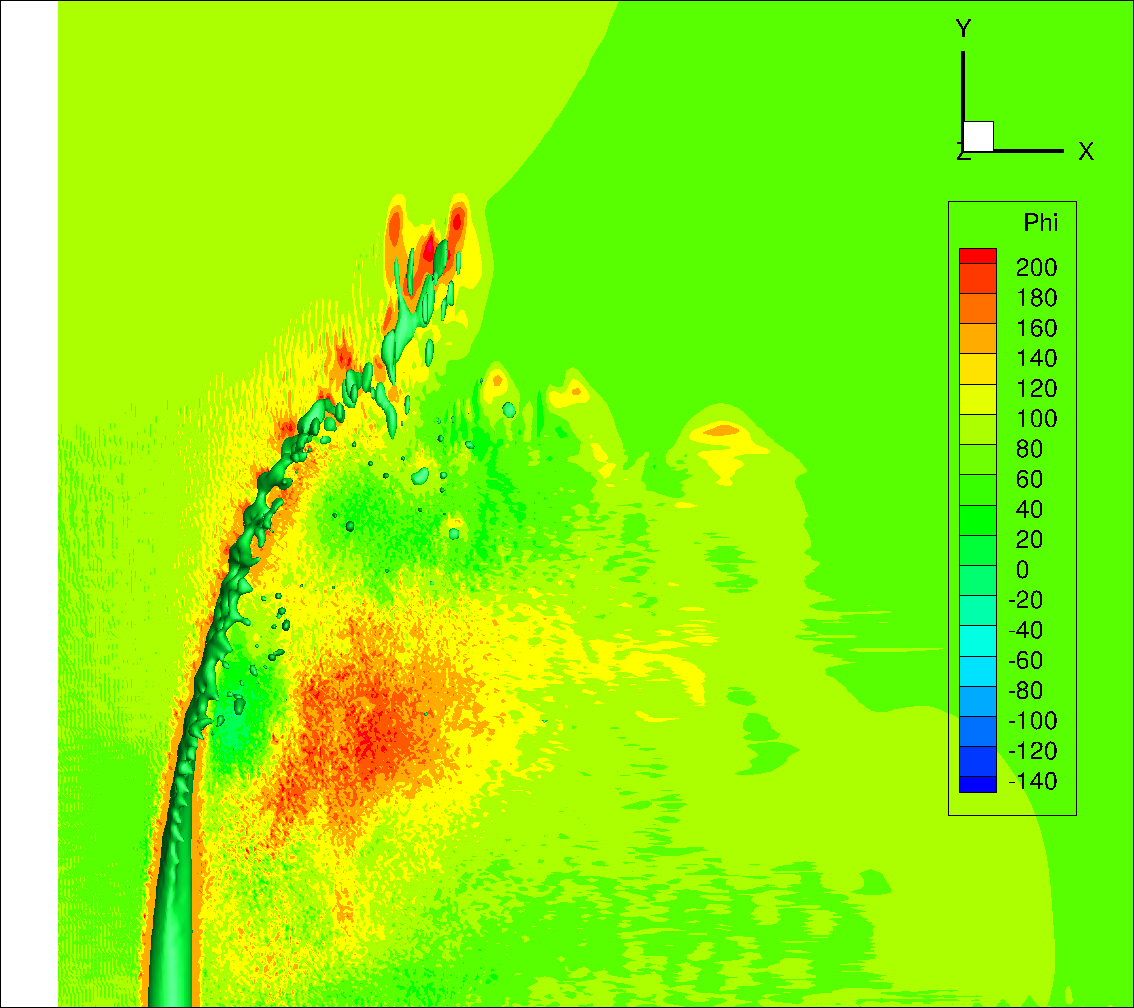} \caption{} 
 \label{fig:jet_con_yz} 
\end{subfigure}\\
\begin{subfigure}{0.32 \linewidth}
\includegraphics[width=\textwidth]{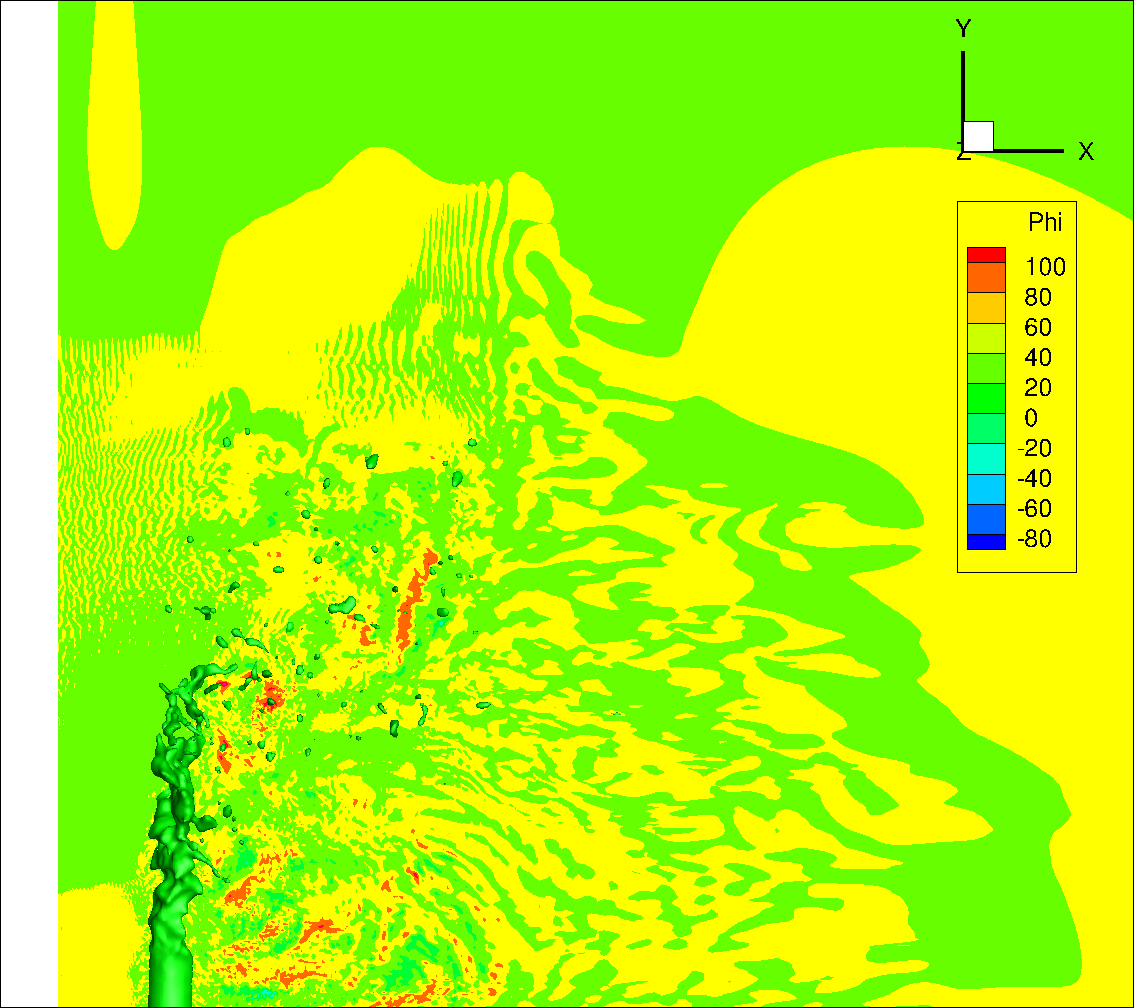} \caption{} 
 \label{fig:jet_con_yz} 
\end{subfigure}
\begin{subfigure}{0.32 \linewidth}
\includegraphics[width=\textwidth]{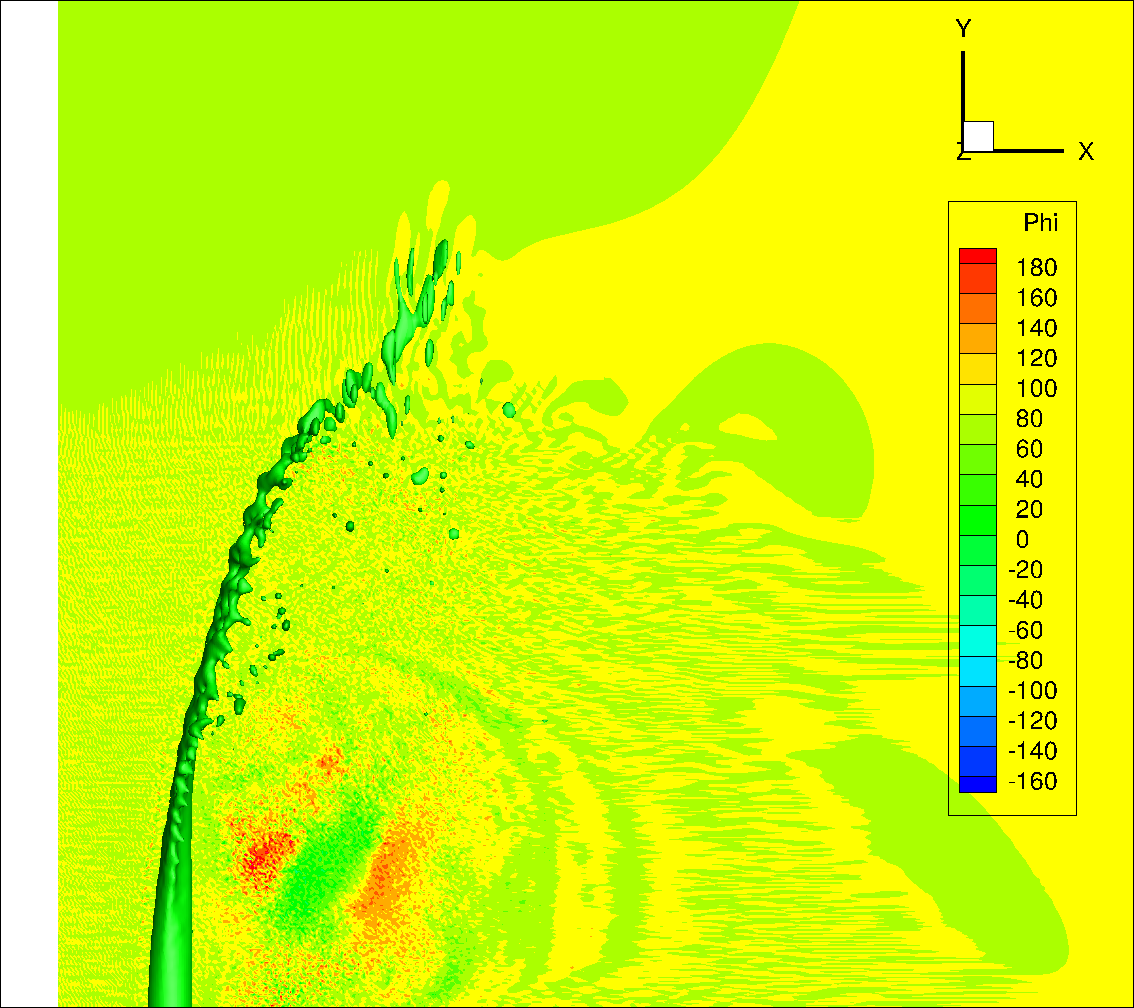} \caption{} 
 \label{fig:jet_con_yz} 
\end{subfigure}
\caption{Results of jet in cross-flow simulation of Case 1 using Equations \ref{NS} and \ref{mom_con} for momentum transport. An $x-y$ view of (a) $u$ from Simuation 1 using Equation \ref{NS} (b) $u$ from Simuation 2 using Equation \ref{mom_con} (c) $v$ from Simuation 1 using Equation \ref{NS} (d) $v$ from Simuation 2 using Equation \ref{mom_con}(e) $w$ from Simuation 1 using Equation \ref{NS} (f) $w$ from Simuation 2 using Equation \ref{mom_con}.}
\label{fig:jet_case_1_coarse_vels}
\end{figure}

In Figure \ref{fig:jet_case_1_coarse_vels}, we show velocity contours in the $x-y$ mid-plane at $t=2.74\times{10}^{-4}s$. A qualitative comparison with \cite{Li2012} (this density ratio was not explored by them) reveals that Simuation 2 is producing much more realistic results.

As we increase the resolution for both the nonconservative and conservative methods (Simuations 3, 4 respectively), solving the nonconservative form of Navier-Stokes (Equation \ref{NS}) in Simuation 3 is no longer robust and the velocity values grow indefinitely around time $t=1\times{10}^{-4}s$. This is due to the inconsistency between this form of the momentum transport equation and the phase field advection equation. Contrarily, Simulation 4 runs stably, avoiding any robustness issues. It is thereby clear that the discrete mass-momentum consistency and conservation of kinetic energy for this form of the momentum transport equation increases the robustness of our diffuse interface method.

\begin{figure}
\centering
\begin{subfigure}{0.49 \linewidth}
\includegraphics[width=\textwidth]{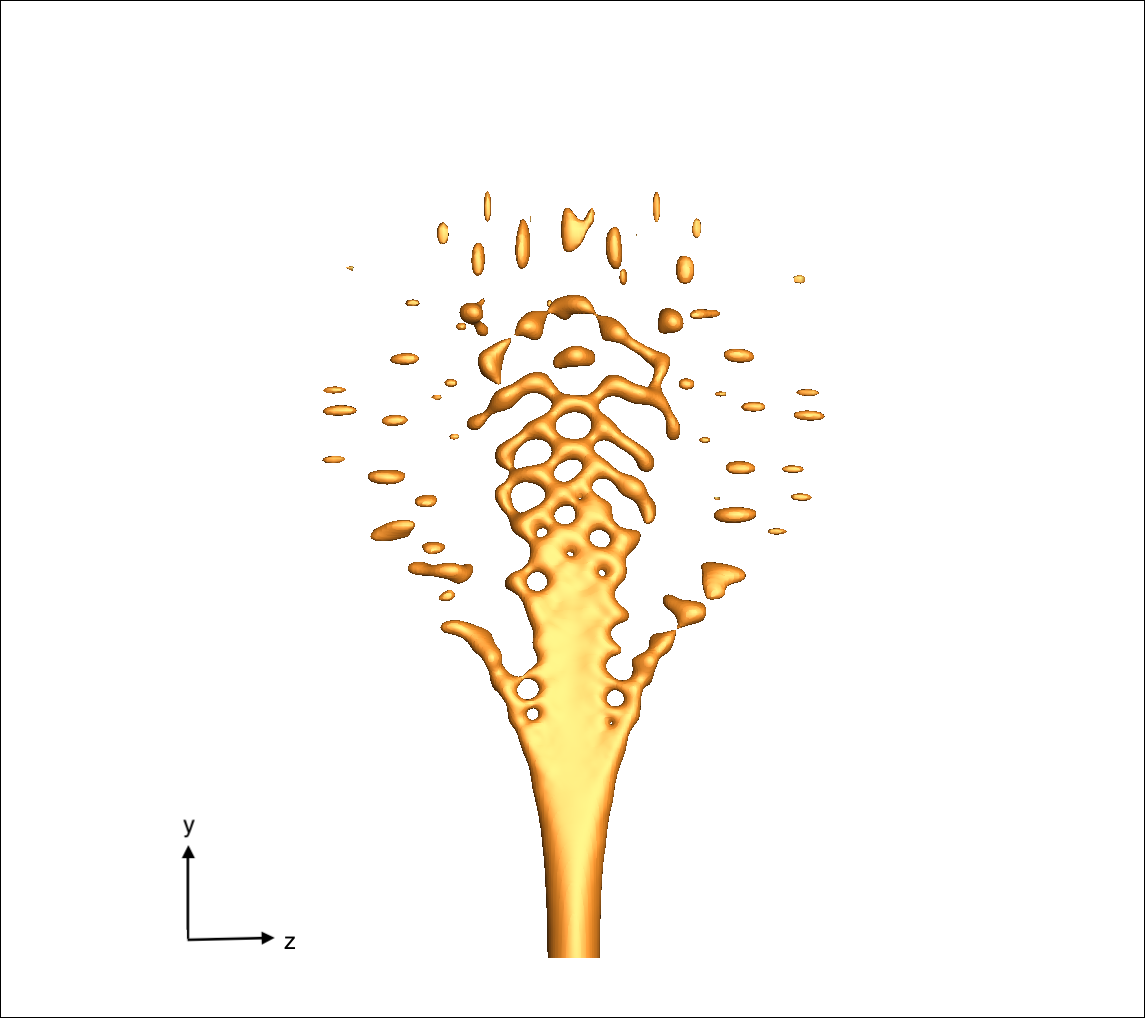} \caption{} 
 \label{fig:jet_hdr_yz}
\end{subfigure}
\begin{subfigure}{0.49 \linewidth}
\includegraphics[width=\textwidth]{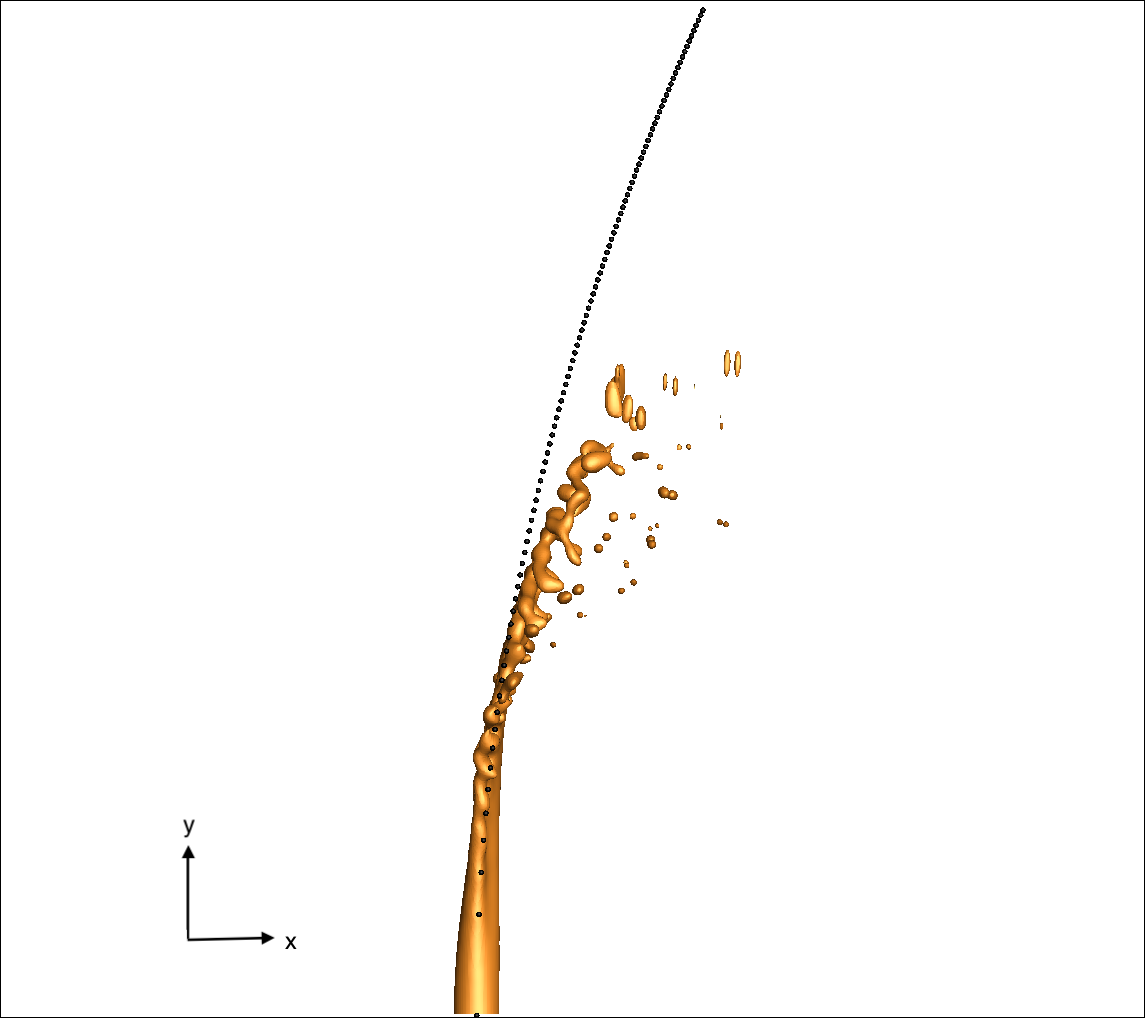} \caption{} 
 \label{fig:jet_hdr_xy} 
\end{subfigure}
\caption{Results of jet in cross-flow simulation of Case 2 using Equation \ref{mom_con} for momentum transport (Simulation 6). Zoom-ins on the jet profile with (a) $y-z$ view (b) $x-y$ view along with the experimental jet trajectory.}
\label{fig:jet_hdr}
\end{figure}

\begin{figure}
\centering
\begin{subfigure}{0.32 \linewidth}
\includegraphics[width=\textwidth]{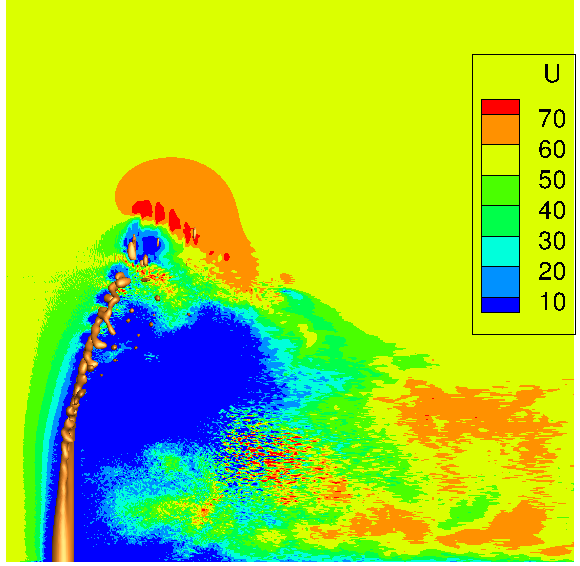} \caption{} 
 \label{fig:jet_hdr_U}
\end{subfigure}
\begin{subfigure}{0.32 \linewidth}
\includegraphics[width=\textwidth]{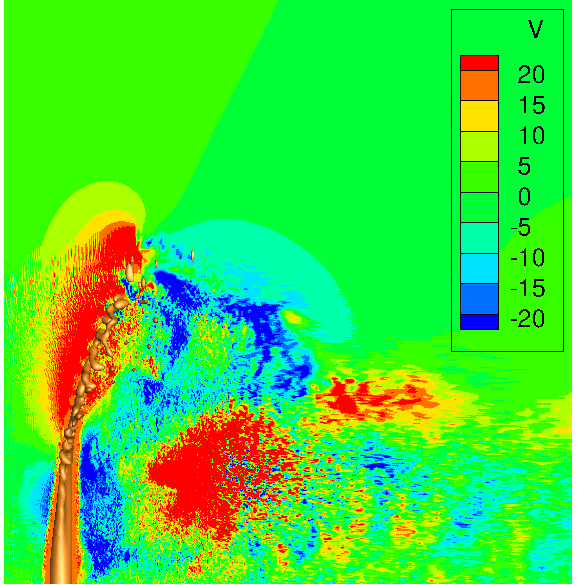} \caption{} 
 \label{fig:jet_hdr_V} 
\end{subfigure}
\begin{subfigure}{0.32 \linewidth}
\includegraphics[width=\textwidth]{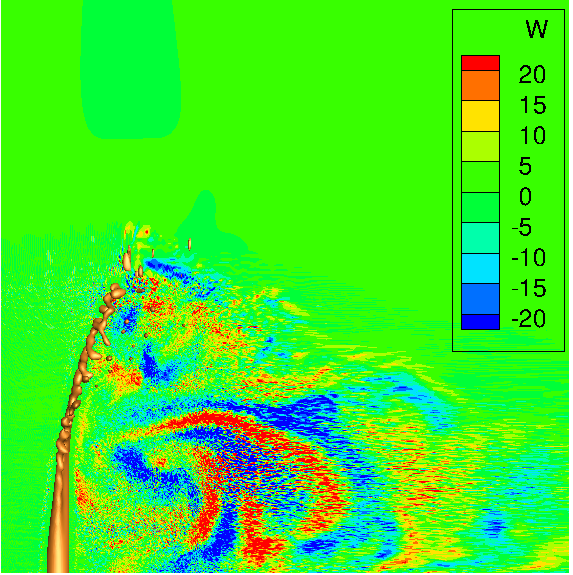} \caption{} 
 \label{fig:jet_hdr_W} 
\end{subfigure}
\caption{Results of jet in cross-flow simulation of Case 2 using Equation \ref{mom_con} for momentum transport (Simulation 6). Displayed are the contour plots on the $x-y$ mid-plane for (a) $u$ (b) $v$ (c) $w$.}
\label{fig:jet_hdr_vels}
\end{figure}

The advantage of using the consistent and conservative formulation for momentum transport is further substantiated as we increase the density ratio and simulate Case 2. Simulations 5 and 6 reveal that while the nonconservative momentum transport equation (Equation \ref{NS}) results in blow-up of velocity magnitudes, our modified momentum transport equation (Equation \ref{mom_con}) captures the jet in a physical manner (with reference to \cite{Sallam2004,Li2012}) even with a relatively coarse mesh for diffuse interface standards (24 cells across jet diameter). Figure \ref{fig:jet_hdr} shows the jet at $t=6\times{10}^{-4}s$ where it has bent significantly and starting to reach a fully developed state. Using the
correlation proposed by \cite{Wu1997} and the adjusted drag coefficient by \cite{Sallam2004}, the experimental jet trajectory is included in Figure \ref{fig:jet_hdr_xy}, showing a good agreement at this resolution.
In Figure \ref{fig:jet_hdr_vels}, the velocity contour values on the $x-y$ mid-plane are depicted for simulation 6. By comparing against the uniform grid CLSVOF results from \cite{Li2012}, which had 20 cells across the jet diameter, the accuracy of our results seem to be promising, especially considering the difference in cost levels between our phase field method and CLSVOF solvers. One can envision that with higher resolution, phase field can possibly yield
more accurate results than those CLSVOF simulations with less cost. More refined calculations can shed
light on this matter.
\section{Proposed free energy based surface tension force}
\label{sec:EBCSF}
Capillary effects play a significant role in many industrial and atmospheric applications such as atomization of liquid jets, ink-jet printers, coating processes and breaking waves. In fact, surface tension is a primary factor in the dynamics of ubiquitous events such as coalescence, break-up, film retraction and instabilities. Such events can lead to topological features such as fingers, crowns, droplets or bubbles. Hence, accurate implementation of surface tension forces is an essential requirement for two-phase flow solvers. Miscalculation of these forces can not only lead to unphysical predictions and failure to correctly capture topological features, but also artificial effects such as spurious currents that can in practice have detrimental effects on the kinetic energy or other integral quantities one may be interested in. 

The need for accurate representation of surface tension has been recognized by researchers. This is reflected by numerous approaches for surface tension modeling within literature. Broadly, surface tension models can be split into integral and volumetric formulations. In integral formulations, surface tension is numerically implemented as a tangential stress at the interface, conserving momentum by default. Continuous surface stress models belong to this category of surface tension formulations. On the other hand, surface tension is implemented as a body force in volumetric formulations (directly in the momentum equation or indirectly as a pressure jump). If accurate, these formulations conveniently allow for discrete balance of pressure and surface tension. Well-balanced integral formulations are much harder to obtain. Very recently, \cite{moataz2018} proposed a two-dimensional algorithm in the context of level-set methods with this property. In general, however, volumetric formulations readily allow for the discrete balance between pressure gradients and capillary forces, preventing the generation of spurious currents, and rendering them much more popular than integral formulations despite failing to conserve total momentum of the system. 

In \cite{Popinet2018}, it was shown that all popular surface tension models are discretely equivalent to the continuous surface force method(CSF) where surface tension is applied as a body force given by 
\begin{equation}
\vec{{ F }_{ ST }}  =\sigma \kappa { { \delta  }_{ \epsilon }\vec{n}  }.
\label{CSF_form}
\end{equation}
In Equation \ref{CSF_form}, ${\delta}_{\epsilon}$ is a numerical Dirac delta function specific to each method. The accuracy of all these methods depends on the accuracy of computing curvature, and their primary goal is to correctly predict the pressure jump across the interface. While the specific manner in which the right hand side terms in Equation \ref{CSF_form} are computed varies, the CSF form can be applied to different one-fluid methods such as volume of fluid, level set and phase field.

In this section, we put forward a different paradigm for computing surface tension forces, and apply it to our selected phase field model. This method for computing surface tension forces utilizes a numerically defined surface energy in a manner consistent with discrete energy conservation principles. Physically speaking, there is work associated with displacing a fluid interface. If surface tension opposes the displacement, then kinetic energy is stored as surface energy. Conversely, if surface tension is pulling the interface in the same direction as its movement, then some of the surface energy is converted to kinetic energy. We use this principle to compute surface tension forces by first defining a numerical surface energy and then computing the surface tension force from the rate of change of surface energy with respect to interface displacement. This novel approach is applied here in the framework of the phase field method given by Equation \ref{phitrans}. The result is a free energy based surface tension force that does not require curvature calculation and reduces spurious currents while preserving the accuracy of the coupled two-phase flow solver. We note that while we present this approach in this framework, the idea of utilizing a defined discrete surface energy to compute surface tension forces can be applied to other interface-capturing schemes likes VOF and level set as well. Similarly, we expect reduction of spurious currents and better total energy conservation when this paradigm is utilized in those frameworks.

In the following, we first discuss the properties a discrete surface energy must have and explain how this idea has been historically used in the context of the Cahn-Hilliard phase field method. Next, we introduce how this idea can be generalized to any interface-capturing method and apply it to our phase field method. We examine the accuracy and advantages of our proposed surface tension force calculation technique via two numerical tests, demonstrating a reduction in spurious currents and improvement in accuracy.

\subsection{Discrete surface energy}
Surface energy is the excess energy at the surface of a material compared to the bulk. Namely, in order to create a surface, molecular bonds within the bulk which are energetically favorable have to be disrupted by some work. That work is stored as surface free energy. Surface tension, $\sigma$, is defined as the amount of work per unit area required to stretch the interface which is equivalent to the amount of surface energy stored per unit area. For any line segment on the surface, $\sigma$ is also the tensile force per unit length exerted tangential to the surface and in the normal direction of the line. Consider a bounded domain $\Omega$ containing a surface $S$. Surface energy density ($\sigma$) can be converted to a volumetric surface energy density via the surface Dirac delta function (${\delta}_{s}$) that is non-zero only on the interface. The total surface energy is then given by
\begin{equation}
{E}_{s}=\oint _{ S }{ \sigma dS }=\int _{ \Omega  }{\sigma{\delta}_{s}(\vec{x}-\vec{{x}_{s}})dV}.
\label{E_s_def}
\end{equation}

For one-fluid models, numerical techniques for including surface tension forces all effectively smoothen the force as implied by Equation \ref{CSF_form}. This is expected for diffuse interface approaches, but also turns out to be the case for sharp interface approaches as no mesh can resolve the infinitesimally thin interface between phases. Following the same logic, numerical volumetric surface energy density, ${\rho}_{s}$, which we will refer to as free energy density for convenience throughout this paper, can be represented through a numerical Dirac delta function, 
\begin{equation}
{\rho}_{s}=\sigma{\delta}_{e}(\vec{x}-\vec{{x}_{s}}).
\label{rho_s_def}
\end{equation}

A numerical model for ${\rho}_{s}$ must satisfy the following properties:
\begin{itemize}
\item In the absence of interfaces, ${\rho}_{s}=0$ everywhere and by definition the free energy stored on the interface of two phases is positive, thus ${\rho}_{s}\ge0$.
\item One of the practical implications of mesh refinement ($\Delta\rightarrow0$) is the reduction of the width of the numerical Dirac delta function denoted by $w({\delta}_{e})$. As the mesh is refined, the volumetric surface energy should converge towards the analytical value for free surface energy,
\begin{equation}
\lim_{\Delta\rightarrow0}{\int_{\Omega}{{\rho}_{s}dV}}={E}_{s}=\oint _{ S }{ \sigma dS }.
\label{energy_limit}
\end{equation}
\item Equation \ref{energy_limit} is valid for any control volume. Imagine an arbitrary surface element contained in a control volume obtained from sufficiently extruding the surface element along the normal vector in both directions (based on the numerical width of the interface, $w({\delta}_{e})$). Then, Equation \ref{energy_limit} would give
\begin{equation}
\lim_{\Delta\rightarrow0}{\int_{-w({\delta}_{e})}^{w({\delta}_{e})}{\rho}_{s}\vec{d{x}_{n}}}=\sigma,
\label{energy_limit_cont}
\end{equation}
where $\vec{d{x}_{n}}$ is a vector in the normal direction to the interface. 

In all one-fluid interface capturing schemes, a phase indicator function is advected, either geometrically or algebraically, from which the interface is implicitly obtained. In all phase field methods, the phase field denoted by $\phi$ is a smoothly varying scalar which assumes constant preset values at pure phases. In such methods, the interface is thickened artificially in the direction normal to the surface and Equation \ref{energy_limit_cont} becomes
\begin{equation}
\lim_{\Delta\rightarrow0}{\int_{-\infty}^{\infty}{\rho}_{s}(\phi)\vec{d{x}_{n}}}=\sigma.
\label{energy_limit_pf}
\end{equation}
This equation is used in phase field methods when calculating the values for the constants involved in computing the free energy density (${\rho}_{s}$).
\end{itemize}
\subsection{Energy-based surface tension force for Cahn-Hilliard}
\label{sec:ebcsf_ch}
The PDE in Cahn-Hilliard drives towards a state of minimum free surface energy, where the free surface energy is defined in Equation \ref{CH_energy_defn}. Based on Equation \ref{CH_energy_defn}, it is clear that for Cahn-Hilliard, ${\rho}_{s}(\phi)={ \epsilon  }^{ 2 }{ \left| \nabla \phi  \right|  }^{ 2 }+W(\phi)$. Surface energy is thus defined as
\begin{equation}
    SE=\int_{\Omega}{{\rho}_{s}dV}.
    \label{SE_def_phase_field}
\end{equation}

Equation \ref{Cahn_Hilliard} can be written in the form
\begin{equation}
\frac { \partial \phi  }{ \partial t } + \nabla\cdot(\vec { u }\phi)={ \nabla  }^{ 2 }{\mu}_{s},
\label{Cahn_Hilliard_ST}
\end{equation}
where ${\mu}_{s}$ is the functional derivative of the free surface energy with respect to $\phi$, also known as the chemical potential,
\begin{equation}
{\mu}_{s}(\phi)=\frac { \delta { \rho  }_{ s } }{ \delta \phi  } ={ \epsilon  }^{ 2 }{ \nabla  }^{ 2 }\phi -W'(\phi ).
\end{equation}
As explained in the introduction, surface tension force can be defined as the rate of change of free surface energy with respect to interface displacement. Having such surface tension forcing would automatically enforce the increase in surface energy due to phase advection to balance the work done against surface tension (kinetic energy loss) during interface displacement. Phase field methods are suitable candidates for implementing this idea as the free surface energy is often a well-defined function of $\phi$ and its derivatives. Indeed, \cite{Jacqmin1999} employed this paradigm for the Cahn-Hilliard equation by arguing that for arbitrary interface configurations and flow fields, there must be a diffuse force exerted by the fluid such that the change in kinetic energy is always opposite to the change in surface free energy. In the continuous limit, for an incompressible flow, the rate of change of free energy due to convection at any point in the domain is
\begin{equation}
{(\frac{\partial {\rho}_{s}}{\partial t})}_{conv}=\frac{\delta {\rho}_{s}}{\delta \phi}{(\frac{\partial\phi}{\partial t})}_{conv}={\mu}_{s}{(\frac{\partial\phi}{\partial t})}_{conv}=-{\mu}_{s}\nabla\cdot(\vec { u }\phi)=-{\mu}_{s}\vec{ u }\cdot(\nabla\phi).
\label{Cahn-Hilliard_energy_rate_transfer}
\end{equation}
On the other hand, the change in kinetic energy in the form of work to overcome capillary forces, represented by $\vec{{F}_{ST}}$, is $\vec{u}\cdot\vec{{F}_{ST}}$. Therefore, based on Equation \ref{Cahn-Hilliard_energy_rate_transfer}, they concluded that for the change in kinetic energy to balance the change in surface energy, 
\begin{equation}
{F}_{ST}={\mu}_{s}\nabla\phi. 
\label{eqn:CH_ST_force}
\end{equation}
This surface tension forcing appears on the right hand side of the momentum equation, as can be seen in Equations \ref{NS}, \ref{momentum_generic_one_fluid_conservative} and \ref{mom_con}.

For now, let's assume that the two phases have the same density, or similar to \cite{Jacqmin1999}, the density variations are small such that Boussinesq approximation can be used. Equations \ref{NS} and \ref{mom_con} are equivalent with this assumption. If we multiply Equation \ref{Cahn_Hilliard_ST} by ${\mu}_{s}$ and Equation \ref{NS} by $\vec{u}$, sum and integrate throughout $\Omega$, we find the rate of change of total energy, given by
\begin{equation}
\frac { \partial (KE+SE) }{ \partial t } =\int _{ \Omega  }{[ -\mu (\nabla {\vec{u} }\cdot\nabla {\vec{ u }})-\kappa (\nabla { { \mu  }_{ s } }\cdot\nabla { \mu  }_{ s })] dV },
\label{total_energy_CH}
\end{equation}
which is clearly negative. At the discrete level, \cite{Jacqmin1999} first defined a discrete surface energy and kinetic energy norm for a staggered uniform mesh configuration, and then showed that the above properties would still hold by using central differences for spatial discretization. Specifically, when displacing an interface, the increase in discrete surface energy due to phase advection would balance the discrete kinetic energy spent on overcoming the surface tension force. Moreover, a discrete version of Equation \ref{total_energy_CH} was found in which the rate of change of discrete total energy in a bounded domain would be always non-positive, resulting in the rapid elimination of spurious currents. From this, it can be interpreted that while being robust, using this surface tension forcing along with the Cahn-Hilliard equation results in a dissipative method, in the discrete and continuous sense. Later, \cite{Shen_Yang} extended this work to non-unity density ratios by modifying the momentum equation.

As explained in Section \ref{sec:Intro}, phase field models based on the Cahn-Hilliard equation suffer from inherent issues that can hinder their performance in realistic applications such as turbulent two-phase flows. 
\subsection{Energy-based surface tension calculation for current phase field}
\label{sec:EBCSF}
In Sections \ref{sec:phase_field} we introduced a phase field model (Equation \ref{phitrans}) that does not suffer from the aforementioned inherent problems of the Cahn-Hilliard equation. We also provided a modified momentum transport equation (Equation \ref{mom_con}) that in the absence of capillary and viscous effects, discretely conserves momentum and kinetic energy. Phase field methods are amenable to defining interfacial quantities such as surface energy, as they readily transform all surface quantites to volumetric functions in space. In this section, after defining a discrete surface energy functional for our phase field method, we apply the physical principle explained above to compute surface tension force as the discrete rate of change of surface energy to interface displacement.

The right-hand side of our phase field introduced in Equation \ref{phitrans}, reproduced here for convenience,
\begin{equation}
    \frac { \partial \phi  }{ \partial t } +\nabla \cdot \left(\vec{ u } \phi \right)=\nabla \cdot \left[\gamma \left(\epsilon \nabla \phi -\phi \left(1-\phi \right)\frac { { \nabla \phi  }  }{ \left| { \nabla \phi  }  \right|  }  \right) \right],
    \label{phitrans2}
\end{equation}
is not a gradient flow to any known energy (right-hand side does not strictly decrease surface energy). However, just like the Cahn-Hilliard equation, the equilibrium profile from Equation \ref{phitrans2} is a hyperbolic-tangent profile and as a result, the Ginzburg-Landau-Wilson free energy functional can be used to obtain an energy that we know is minimized at equilibrium. As a matter of fact, we have numerically found that when $\vec{u}=0$, the surface energy functional given by 
\begin{equation}
SE=\int _{ \Omega  }{{\rho}_{se}dV}=\int _{ \Omega  }{ \frac{3\sigma}{\epsilon}\left[{ \epsilon  }^{ 2 }{ \left| \nabla \phi  \right|  }^{ 2 }+{({\phi}^{2}-\phi)}^{2}\right]dV },
\label{SE_def_our_pf}
\end{equation}
is strictly decreased by Equation \ref{phitrans2} when applied to any tangent hyperbolic profile (in the form of ${\phi}_{0}=(1+tanh(\vec{x}/{2\epsilon}))/2$, where $\epsilon$ is arbitrary), as it sharpens or diffuses the interface to its correct thickness, enforced by $\epsilon$. The potential is then
\begin{equation}
{\mu}_{se}=\frac{\delta SE}{\delta \phi}= \frac{6\sigma}{\epsilon}\left[-{ \epsilon  }^{ 2 }{\nabla}^{2}\phi+\phi(\phi-1)(2\phi-1)\right].
\label{pot_def}
\end{equation}
 Recall that surface tension force can be defined as the rate of change of free surface energy with respect to interface displacement. With this definition, any rise in surface energy is balanced by a corresponding dip in kinetic energy and vice versa. There is no true interface in phase field methods. Instead, the phase field variable, $\phi$ is advected in the domain. The rate at which the discrete surface energy in the domain changes as $\phi$ and its higher derivatives are altered in response to advection of $\phi$, is the discrete value for the surface tension force. From Equation \ref{phitrans2} the rate at which $\phi$ varies due to advection is ${(\partial\phi/\partial t)}_{conv}=-\nabla\cdot(\vec{u}{\phi})=-\vec{u}\cdot\nabla\phi$. The rate of change of discrete surface energy with respect to time due to phase field advection is $\partial {\rho}_{s}/\partial t=(\delta {\rho}_{s}/\delta \phi){(\partial\phi/\partial t)}_{conv}$. The rate at which $\phi$ is displaced due to advection is given by $-\vec{u}$. As such, surface tension force, defined as the rate of change of surface energy per displacement of $\phi$ is given by $(\partial{\rho}_{s}/\partial t)/(-u)$, 
\begin{equation}
\vec{{F}_{ST}}=(\delta {\rho}_{s}/\delta \phi)\nabla\phi={\mu}_{s}\nabla\phi. 
\label{ebcsf_formula}
\end{equation}
Notice that unlike \cite{Jacqmin1999}, we did not explicitly attempt to balance the increase in kinetic energy with the decrease in surface energy and vice versa. Instead, we practiced the general paradigm, that locally, on any point on the interface, surface tension force is equal to the rate of energy exchange to/from surface energy per unit displacement of the interface.

Consider a two-phase system with ${\rho}_{1}={\rho}_{2}$, for which Equations \ref{NS} and \ref{mom_con} are equivalent. After substituting the surface tension force ($\vec{{ F }_{ ST }}={\mu}_{s}\nabla \phi$) in either of these these equations, say Equation \ref{mom_con}, summing the product of Equation \ref{phitrans2} with the potential and Equation \ref{mom_con} with $\vec{u}$, and then integrating gives the rate of change of total energy to be
\begin{equation}
\frac { \partial (KE+SE) }{ \partial t } =\int _{ \Omega  }{ \left\{-\mu (\nabla \vec{u}.\nabla \vec{u})+{\mu}_{s}\gamma \nabla \cdot \left  [\epsilon \nabla \phi -\phi (1-\phi ) \left ( \frac { { \nabla  }\phi  }{ \left| { \nabla  }\phi  \right|  } \right ) \right ]\right\}dV }.
\label{total_energy_balance_di}
\end{equation}
We emphasize that on the PDE level, Equation \ref{total_energy_balance_di} implies the balance between the increase of $SE$ due to advection, ${\mu}_{s}\nabla\cdot(\vec{u}\phi)$, and work done to oppose the surface tension force, $\vec{u}\cdot{F}_{ST}=\vec{u}\cdot({\mu}_{s}\nabla \phi)$. We maintain this property discretely by using central differences on our staggered grid configuration. The first term on the right-hand-side of Equation \ref{total_energy_balance_di} represents viscous dissipation and is always negative. However, as explained before, the second term which represents the rate of change of surface energy due to the compression and diffusion terms in the right-hand-side of Equation \ref{phitrans2} is not necessarily negative. While this does not guarantee elimination of spurious currents, it demonstrates that our chosen phase field method is less dissipative compared to Cahn-Hilliard. Indeed however, in the following we numerically show that compared to the original CSF model, the new energy-based formulation significantly reduces spurious currents while also improving accuracy of surface tension force calculation.
\subsubsection{Spurious Currents}
\label{sec:spur_cur}
This test case is a standard benchmark for two-phase flow solvers\citep{Williams1998,Francois2006,Herrmann2008}. A 2D drop with diameter $D=0.4$ is placed in a $1\times1$ box with slip boundary conditions and zero initial flow. The drop and surrounding fluid have equal densities and viscosities. Theoretically, no flow should be generated, and the drop and its background should remain unchanged. 

\begin{figure}
\centering
\includegraphics[width=0.75\textwidth]{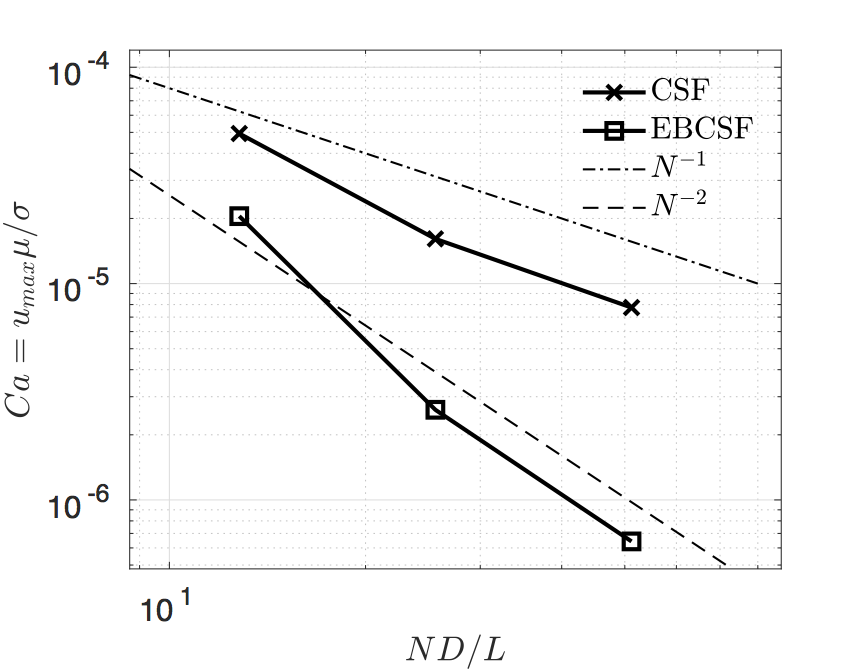}
\caption{Spurious currents for the curvature-based CSF formulation versus the new curvature-free method, which we denote as energy-based continuous surface force, EBCSF.}
\label{fig:spurious_currents_ebcsf}
\end{figure}

The dimensionless parameter characterizing this problem is the Laplace number, $La=\sigma \rho D / { \mu }^{ 2 }$. We set $La=12000$ for our comparison between the original continuum surface force (CSF) model and our energy-based model continuum surface force model (Equation \ref{ebcsf_formula}), which we denote by EBCSF. Both the CFS and EBCSF model are balanced-force methods\citep{Francois2006}. As such, if the surface tension forces were computed exactly, both would generate no spurious currents, as they allow for the discrete balance between pressure gradients and surface tension forces. Therefore, any sustained flow in the numerical solution is due to errors in computing the surface tension force. We will examine the magnitude of the largest velocity in the domain at non dimensional time $t^*=\sigma t/( \mu D ) =250$, which is a large enough time for the surface tension and viscous forces to have physically balanced each other and for spurious currents to be fully developed. To obtain optimal convergence using DI, we set $\epsilon={\Delta x}^{2/3}/\sqrt[3]{32}$. In Figure \ref{fig:spurious_currents_ebcsf}, the magnitude of $Ca={u}_{max}\mu/\sigma$ is plotted at various levels of refinement. It is clear that EBCSF  reduces spurious currents considerably compared to CSF. Moreover, the convergence rate of EBCSF seems to be higher than CSF as well.
\subsubsection{Standing Wave}
\label{sec:standing_wave}
\begin{figure} 
\centering
\begin{subfigure}{0.49 \linewidth}
\includegraphics[width=0.95\textwidth]{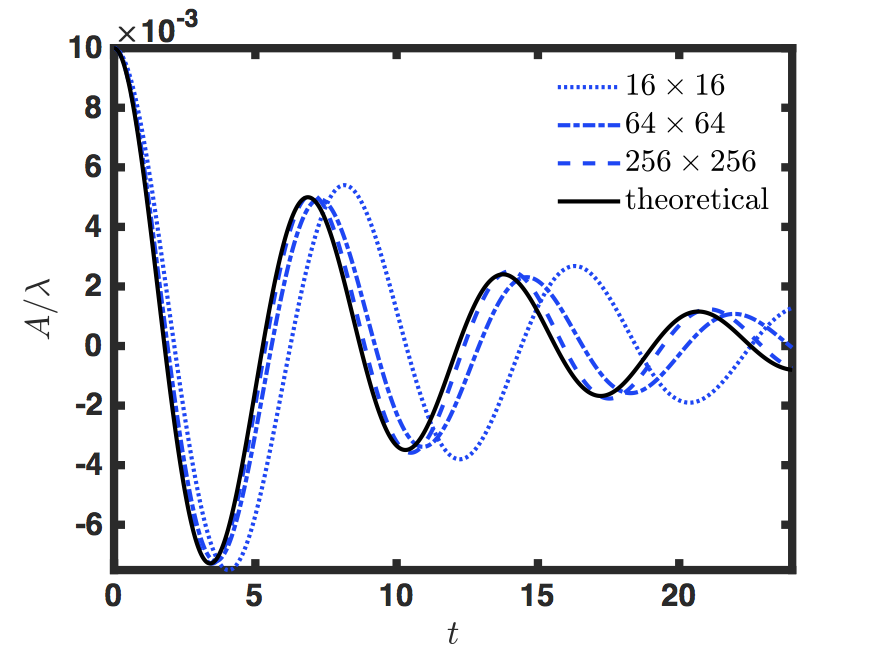} 
   \caption{}
   \label{fig:ebscf_sw} 
\end{subfigure}
\begin{subfigure}{0.49 \linewidth}
\includegraphics[width=0.95\textwidth]
{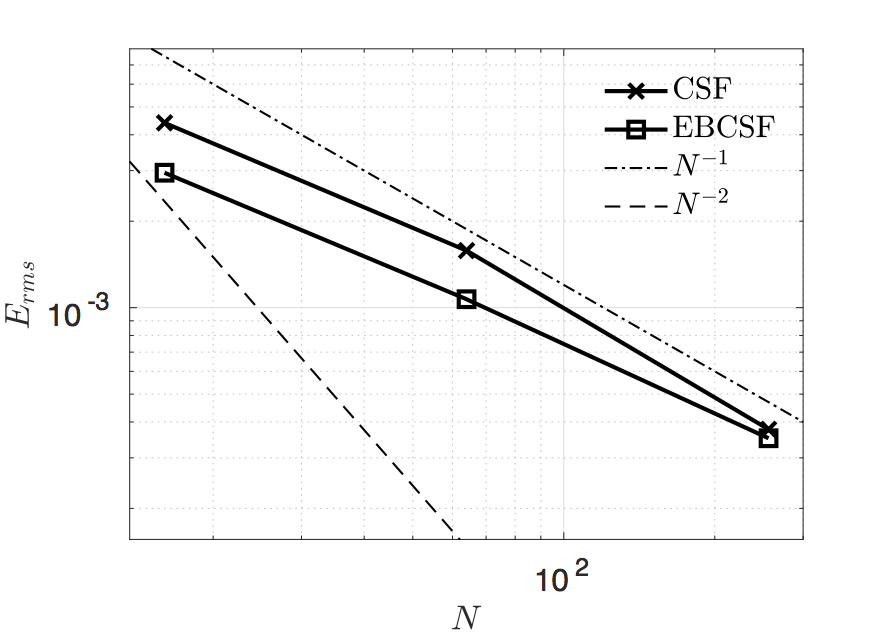} 
   \caption{}
  \label{fig:surface_wave_di}  
   \end{subfigure}
    \caption{Amplitude of the standing wave at $x=\Delta x/2$ using the EBCSF surface tension force method against the theoretical solution (a), r.m.s error values for interface location against resolution for standing wave simulations using CSF and EBCSF (b).}
\end{figure}

In this 2D test case \citep{Popinet1999,Gueyffier1999,Gerlach2006,Herrmann2008}, by examining surface oscillations of a standing wave, one can assess the accuracy of a two-phase flow solver in capturing the interaction between inertial, capillary and viscous forces. Here, we utilize this test to verify the accuracy of EBCSF in computing surface tension forces. A single small amplitude wave with wavelength $\lambda=2\pi$ is placed between two immiscible fluids in a $[0,2\pi]\times[0,2\pi]$ domain. The initial perturbation amplitude is ${A}_{0}=0.01\lambda$ and ${y}_{0}=\pi$. The initial condition for the wave height is
\begin{equation}
{ h }_{ wave }(x,t=0)=y-{ y }_{ 0 }+{ A }_{ 0 }cos(x-\Delta x/2).
\end{equation}
Boundary conditions are periodic in the $x$ direction and slip boundary conditions are used for the top and bottom walls. If the two fluids have equal kinematic viscosity, $\nu$, then the amplitude of the wave is given analytically(\cite{Prosperetti1981}):
\begin{equation}
{ A }_{ ex }=\frac { 4(1-4\beta ){ \nu  }^{ 2 } }{ 8(1-4\beta ){ \nu  }^{ 2 }+{ \omega  }_{ 0 }^{ 2 } } { A }_{ 0 }erfc\sqrt { \nu t } +\sum _{ i=1 }^{ 4 }{ \frac { { z }_{ i } }{ { Z }_{ i } } (\frac { { \omega  }_{ 0 }^{ 2 }{ A }_{ 0 } }{ { z }_{ i }^{ 2 }-\nu  } )exp[({ z }_{ i }^{ 2 }-\nu )t]erfc({ z }_{ i }\sqrt { t } ) },
\label{eqn:prosperetti}
\end{equation}
where ${z}_{i}$ are the roots of
\begin{equation}
{ z }^{ 4 }-4\beta \sqrt { \nu  } { z }^{ 3 }+2(1-6\beta ){ \nu z }^{ 2 }+4(1-3\beta ){ \nu  }^{ 3/2 }z+(1-4\beta ){ \nu  }^{ 2 }+{ \omega  }_{ 0 }^{ 2 }=0.
\end{equation}
Here, we set $\sigma=2$ ${\rho}_{1}={\rho}_{2}=1$, $\nu=0.064720863$ and the time step is $dt=0.003$. At different resolutions, we measure the amplitude of the standing wave at $x=\Delta x/2$. The amplitude of the standing wave for simulations performed with EBCSF are plotted against the theoretical solution in Figure \ref{fig:ebscf_sw}. It is clear that with higher resolution, the solution is converging to the theoretical prediction given by \ref{eqn:prosperetti}. After computing the error with respect to the theoretical solution, we compare the root-mean-squared (r.m.s.) error of the simulation predictions using CSF and EBCSF in Figure \ref{fig:surface_wave_di}. Both surface tension computation schemes give first order convergence rates, while EBCSF proves to be more accurate than CSF.

Lastly, using Equation \ref{mom_con}, if we combine the corrections introduced in Section \ref{sec:consistent_conservative_momentum_transport} with the energy-based surface tension force, we recover Equation \ref{total_energy_balance_di} for non-unity density ratios.

\section{Conclusions}
\label{sec:Conclusions}
In the framework of a conservative and bounded second order phase field method, we have introduced a momentum and kinetic energy conserving momentum transport model. This proposed model is consistent with the phase/mass transport equation, resulting in significant improvement in the robustness, stability and accuracy of two-phase flow simulations at high density ratios and/or high $Re$ numbers. This was confirmed via canonical and realistic problems studied herein. 

Additionally, a general paradigm for numerical evaluation of surface tension forces is presented and implemented on the adopted phase field method. This new surface tension force model does not require computation of curvature values, and instead utilized discrete definition of free surface energy. We numerically demonstrated that this model is an upgrade to the the commonly used CSF model.

All in all, the two main contributions in this work can be adapted to other two-phase flow models. For phase field models, the momentum treatment can be easily adapted, allowing for robust and accurate simulations at high density ratios and turbulent conditions. Moreover, for various interface-capturing methods, after defining a discrete surface energy density, one can take advantage of the surface tension paradigm introduced here to compute surface tension forces. This is expected to improve the conservation of total energy and capillary force predictions.

\begin{appendices}
\section{Non-uniform Cartesian grids}
\label{sec:non_uniform_mesh}
In this section, we first explain how Equation \ref{phitrans} should be discretized such that the boundedness properties of the proposed second order phase field method can be extended to non-uniform grids. Then, we present the discretization strategy for Equation \ref{mom_con} that will allow for consistent and conservative momentum transport on non-uniform grids.
\subsection{Phase field advection}
The thickness of the interface in phase field methods is usually picked to be a globally constant value. In order to be able to resolve the interface everywhere in the domain, the thickness would then be selected based on the coarsest mesh size. This results in either thick interfaces or over-resolved regions. We here propose an alternate strategy for Equation \ref{phitrans}, in which the interface thickness, given by $\epsilon$, varies according to the local mesh. The space-time discretization for forward Euler time-stepping in 2-D for Equation \ref{phitrans} becomes
\begin{multline}
{ \phi  }_{ i,j }^{ k+1 }={ \phi  }_{ i,j }^{ k }+\Delta t(-\frac { { u }_{ i+1/2,j }^{ k }({ \phi  }_{ i+1,j }^{ k }+{ \phi  }_{ i,j }^{ k })/2-{ u }_{ i-1/2,j }^{ k }({ \phi  }_{ i,j }^{ k }+{ \phi  }_{ i-1,j }^{ k })/2 }{ { \Delta x }_{ i,j } } -\\\frac { { v }_{ i,j+1/2 }^{ k }({ \phi  }_{ i,j+1 }^{ k }+{ \phi  }_{ i,j }^{ k })/2-{ v }_{ i,j-1/2 }^{ k }({ \phi  }_{ i,j }^{ k }+{ \phi  }_{ i,j-1 }^{ k })/2 }{ { \Delta y }_{ i,j } } \\ \gamma \frac { { \epsilon  }_{ x,i+1/2,j }({ \phi  }_{ i+1,j }^{ k }-{ \phi  }_{ i,j }^{ k })/{ \Delta x }_{ i+1/2,j }-{ \epsilon  }_{ x,i-1/2,j }({ \phi  }_{ i,j }^{ k }-{ \phi  }_{ i-1,j }^{ k })/{ \Delta x }_{ i-1/2,j } }{ { \Delta x }_{ i,j } } +\\\gamma \frac { { \epsilon  }_{ y,i,j+1/2 }({ \phi  }_{ i,j+1 }^{ k }-{ \phi  }_{ i,j }^{ k })/{ \Delta y }_{ i,j+1/2 }-{ \epsilon  }_{ y,i,j-1/2 }({ \phi  }_{ i,j }^{ k }-{ \phi  }_{ i,j-1 }^{ k })/{ \Delta y }_{ i,j-1/2 } }{ { \Delta y }_{ i,j } } +\\ \gamma \frac { { ({ (\phi  }_{ i+1,j }^{ k }) }^{ 2 }-{ \phi  }_{ i+1,j }^{ k })\hat { { n }^{ k }_{ i+1,j } } -{ { ((\phi  }_{ i-1,j }^{ k }) }^{ 2 }-{ \phi  }_{ i-1,j }^{ k })\hat { { n }^{ k }_{ i-1,j } }  }{ { 2\Delta x }_{ i,j } } +\\\gamma \frac { { ({ (\phi  }_{ i,j+1 }^{ k }) }^{ 2 }-{ \phi  }_{ i,j+1 }^{ k })\hat { { n }^{ k }_{ i,j+1 } } -{ { ((\phi  }_{ i,j-1 }^{ k }) }^{ 2 }-{ \phi  }_{ i,j-1 }^{ k })\hat { { n }^{ k }_{ i,j-1 } }  }{ { 2\Delta y }_{ i,j } } ),
\label{eqn:no_n_assumption_discrete}
\end{multline}
where ${\Delta x}_{i-1/2,j}$ is the node to node distance between node $(i,j)$ and $(i-1,j)$, computed as ${\Delta x}_{i-1/2,j}=({\Delta x}_{i,j}+{\Delta x}_{i-1,j})/2$, and ${\epsilon}_{x,i-1/2,j}$ is the interface thickness in the $x$ direction at the left face of the cell $(i,j)$. Naturally, ${\Delta y}_{i,j-1/2}$ and ${\epsilon}_{x,i,j-1/2}$ are defined in similar manner. The $\epsilon$ vector is thus stored like the velocity vector. Any arbitrary non-uniform Cartesian grid in physical space can be mapped to an isotropic, uniform Cartesian grid in computational space for which ${\Delta x}_{comp}={\Delta y}_{comp}$ are global constants. Moreover, in order to have constant interface resolution in physical space, we need to have constant interface thickness in computational space. Since the thickness and shape of the interface is determined by the balance of the sharpening and diffusive fluxes on the right-hand side of Equation \ref{phitrans}, we compute these fluxes in the computational space. In that computation, ${\epsilon}_{comp}$ is a global constant. Based on Equation \ref{eqn:no_n_assumption_discrete}, this is equivalent to having set ${\epsilon}_{x,i-1/2,j}/{\Delta y}_{i,j-1/2}={\epsilon}_{y,i,j-1/2}/{\Delta y}_{i,j-1/2}={\epsilon}_{comp}/{\Delta x}_{comp}$ a constant value for all $i$ and $j$. Additionally, the normal vector is computed via finite differences on the computational grid. With these conditions, it can then be proven that boundedness for $\phi$ can be maintained as long as 
\begin{equation}
    \frac{{\epsilon}_{comp}}{{\Delta x}_{comp}}\ge\frac { { \gamma  }/{\left|\vec{u}\right|}_{max}+1 }{ 2{ \gamma  }/{\left|\vec{u}\right|}_{max}},
    \label{criteria_nonuniform}
\end{equation}
and the time-step is chosen appropriately based on numerical stability,
\begin{equation}
    \Delta t \le \left[\frac{{\left|\vec{u}\right|}_{max}/\gamma}{{\epsilon}_{comp}/{\Delta x}_{comp}}\right]\frac{min({\Delta x}_{min},{\Delta y}_{min})}{2\left|\vec{u}\right|_{max}},
\end{equation}
where $\left[{(\left|\vec{u}\right|}_{max}/\gamma)/({\epsilon}_{comp}/{\Delta x}_{comp})\right]$ is a constant chosen by the user according to Equation \ref{criteria_nonuniform}.
\subsection{Momentum transport}
Following the work of \cite{Ham2002}, when dealing with non-uniform Cartesian grids, we use volume weighted averaging when computing convective fluxes in Equation \ref{mom_con}. To interpolate a field $\psi$ from cell center to x-face using volume weighted averaging, 
\begin{equation}
    {\tilde{\psi}}_{i-1/2,j}=\frac{{\psi}_{i-1,j}{\Delta x}_{i-1,j}/2+{\psi}_{i,j}{\Delta x}_{i,j}/2}{{\Delta x}_{i-1/2,j}}
\end{equation}
In order to have consistent and kinetic energy conserving momentum transport we use this type of averaging in 
\begin{itemize}
\item interpolating $\rho$ in the temporal terms from nodes onto face cells in Equation \ref{mom_con_index_form} (this interpolated value is used when solving the pressure Poisson equation)
\item interpolating ${U}_{j}$ in Equation \ref{mom_con_index_form} for $j\neq i$
\item interpolating $\phi$ onto face centers when computing $\nabla\cdot(\vec{u}\phi)$
\end{itemize}
\end{appendices}
\section{Acknowledgements}
This work was financially supported by Office of Naval Research (Grant No. 119675) and NASA (Grant No. 127881). Mr. Pedro Milani is acknowledged for his contribution to the development of the energy-based scheme for surface tension forces.
\section{References}

\bibliography{mybibfile}

\end{document}